\documentclass{article}

\usepackage{amsmath,amsfonts,amssymb}
\usepackage{algorithmic}
\usepackage{array}
\usepackage[caption=false,font=normalsize,labelfont=sf,textfont=sf]{subfig}
\usepackage{textcomp}
\usepackage{stfloats}
\usepackage{url}
\usepackage{verbatim}
\usepackage{graphicx}
\usepackage{float}
\usepackage{lipsum}
\usepackage{svg}
\usepackage{hyperref}

\newcommand{\R}{\mathbb{R}}

\newcommand{\cS}{\mathcal{S}}

\newcommand{\cX}{\mathcal{X}}

\newcommand{\cB}{\mathcal{B}}

\newcommand{\cF}{\mathcal{F}}

\newcommand{\conv}[1]{\operatorname{conv}({#1})}

\usepackage{multicol}
\usepackage{tikz}
\usepackage{psfrag}
\usepackage[toc,page]{appendix}
\usetikzlibrary{arrows, automata, quotes, positioning}
\usepackage{epstopdf}
\newtheorem{definition}{Definition}
\newtheorem{theorem}{Theorem} 

\newtheorem{remark}{Remark}
\newtheorem{lemma}{Lemma}
 \newtheorem{assumption}{Assumption}
\newtheorem{example}{Example}
\def\BibTeX{{\rm B\kern-.05em{\sc i\kern-.025em b}\kern-.08em
    T\kern-.1667em\lower.7ex\hbox{E}\kern-.125emX}}
\begin{document}
\title{Provably Safe Finite-Time  Guidance for Marine Vehicles}
\author{Bhawana Singh, Karim Ahmadi Dastgerdi,\\ Nikolaos Athanasopoulos, Wasif Naeem and  Benoit Lecallard\thanks{Bhawana Singh, Karim Ahmadi Dastgerdi, Nikolaos Athanasopoulos and Wasif Naeem are with the School of Electronics, Electrical Engineering and Computer Science, Queen's University Belfast, Belfast, UK. Benoit Lecallard is with the Artemis Technologies Limited, Belfast, UK.  E-mails: \texttt{\{b.singh, k.ahmadidastgerdi, n.athanasopoulos, w.naeem\} @qub.ac.uk, benoit.lecallard@artemistechnologies.co.uk} 
	This study was conducted as part of the Belfast Maritime Consortium UKRI Strength in Places project, ‘Decarbonisation of Maritime Transportation: A return to Commercial Sailing’ led by Artemis Technologies Limited, Project no. 107138.	
 }
	}
\date{}
	


\maketitle

\begin{abstract}
We consider a new control strategy for marine navigation, equipped with finite-time convergence characteristics.   We provide  mathematical guarantees for waypoint reaching and obstacle avoidance for different encounter scenarios, by deriving conditions under which (i) convergence to waypoint and (ii) safe obstacle avoidance is achieved while (iii)  satisfying  input constraints. We propose a predefined-time heading control to enforce ship heading error convergence and waypoint reaching in finite time. Using this as a building block, we develop a provably safe algorithm for safe waypoint navigation  by strategically and automatically introducing intermediate virtual waypoints. Using Imazu problems as benchmarks, we show that the proposed method is better than other existing strategies such as Velocity Obstacle Avoidance  and biased Line-of-Sight methods, in terms of the safe distance between the ship and the obstacles, cross track error, control effort,  waypoint reaching time and ship path length.
\end{abstract}

\section{Introduction}
The demand for advanced maritime transportation has grown steadily due to the increasing reliance  on ocean and waterway transportation for commercial, freight and leisure activities \cite{ozturk2022review}, \cite{campbell2012review}. Due to this rising ship economy, the potential of accidents among sea-going vessels has greatly increased.  In the European Maritime Safety Agency (EMSA)  report, it was mentioned  that between $2014-2021$, $54\%$ of casualties  were caused by collisions and groundings incidents \cite{EMSA}. 

Typically, motion control systems for marine craft comprise of a guidance subsystem and a heading autopilot control subsystem \cite{breivik2008guidance}. The guidance part is responsible for providing desired commands to obtain a kinematic control objective. In general, desired heading commands are known as guidance laws, while the heading autopilot is responsible for calculating required control forces to track the desired guidance commands.

Control of marine systems with high accuracy and timely maneuvering is challenging as the ship dynamics is nonlinear and the surrounding environment comprises of static (map) and dynamic obstacles (e.g. fishery and other ships).
 Several control system techniques exist in the literature: Dynamic positioning (DP) \cite{sorensen2011survey} for oil exploration and pipeline laying aims to maintain the position and heading of a ship working in full actuation mode at a preset waypoint. Several path-following controllers are designed in the literature  with constant speed profile working in underactuated mode considering the full ship model including both dynamics and kinematics \cite{li2008point}. 

To reduce the complexity introduced by the dynamics, researchers considered a path-following problem using only the kinematic model assuming that the speed of the vessel is controlled independently \cite{do2004robust}. This problem is frequently solved using Line-of-Sight (LOS) laws, in particular, proportional and integral LOS laws \cite{fossen2014uniform,fossen2014line,borhaug2008integral, caharija2016integral}. In general, the LOS guidance law is  straightforward to implement,  mimicking an experienced sailor: they  follow a path specified by preset waypoints asymptotically  by considering cross track errors and lookahead distances. However, computation of cross track errors is calculation intensive and estimation or tuning of look-ahead distances is not always simple. In \cite{wang2019hyperbolic}, a hyperbolic tangent LOS based guidance law is designed to follow a specified path in finite-time, whereby a non-smooth analysis is conducted for mathematical guarantees by utilizing  sliding mode control. Other works provide finite and fixed-time convergence of the cross track errors, such as surge heading LOS based guidance \cite{wang2019surge} and predictor based fixed time LOS guidance \cite{wang2023predictor}. An overview of recent advances in LOS based path following algorithms for marine crafts are in \cite{gu2022advances}. 

A common path planning method  is to select a circle of acceptance (COA) around the waypoint and design guidance laws such that the ship is considered to reach the waypoint when it is in the COA, after which the navigation algorithm selects the next waypoint provided in the mission planner \cite{naeem2012colregs}, \cite{healey1993multivariable}. These algorithms are typically validated through experimental results, lacking  guarantees for waypoint reaching and safe collision avoidance.  
In addition, these methods have the limitation that precision path following in tight space maneuvering is not easily obtained. 

The problem of obtaining collision avoidance guarantees in the presence of obstacles is less studied in the LOS  guidance for marine vehicles. However, it is separately investigated by some other different framework, such as reachability analysis \cite{lakhal2022safe} \cite{xiao2023time, wetzlinger2023fully, krasowski2022commonocean, krasowski2021temporal}. These methods typically utilise  optimal control formulation often computationally complex,  requiring numerical solution of the Hamilton-Jacobi Bellman (HJB) equations. 
Recently, to overcome these challenges of computing reachable sets,  researchers use  control barrier functions (CBFs) based safety guarantees \cite{marley2021maneuvering}, \cite{gao2022safety}.  However, a key challenge in this method lies in the choice of  barrier functions.
Some researchers propose model predictive based control methods to deal with collision avoidance \cite{du2021mpc, du2022colregs}, while Machine learning methods have also been used for collision avoidance and mission planning \cite{sarhadi2022survey}. Limitations include lack of safety and stability guarantees. Another classical guidance method is artificial potential field \cite{ge2002dynamic, li2021path} providing collision-free paths, however, with no target reaching guarantees.
In summary, waypoint LOS based guidance research is commonly found in the literature, however, often without safety guarantees (obstacle avoidance), while on the other hand, methods that do provide these can be conservative (CBFs) or complex (HJB). This motivates the development of simplified guidance laws navigating between waypoints and avoiding obstacles with accuracy even in tight maneuvering space, and providing at the same time convergence and safety guarantees. 

In this context, this work considers waypoint navigation and collision avoidance problems for a marine vehicle navigating in the presence of static and dynamic obstacles. We consider dynamic obstacles  having constant heading angle and velocity. Our new LOS-based strategy works in two modes, namely, nominal guidance mode  and collision avoidance mode.

The core development is that the guidance law works in cascade with a predefined-time heading controller that induces finite-time convergence properties. Specifically, the  nominal guidance mode is designed to navigate towards a waypoint (goal). 
In collision avoidance mode, we compute new intermediate virtual waypoints according to the bound of \textit{distance to the closest point of approach} (DCPA) \cite{hu2019multiobjective} that is known as \emph{closest point of approach (CPA)} around the obstacle position. The desired guidance law is next designed for the intermediate virtual waypoint navigation.  To track the desired guidance law in both modes,  a predefined-time control strategy is introduced for heading autopilot design and control Lyapunov function methods induce the control law. Since the equilibrium point of the autopilot error dynamics is predefined-time stable, we are able to provide  guarantees for finite time waypoint convergence. Last, we calculate a virtual unsafe set according to the CPA that includes the obstacle at all times, which we guarantee to be avoided, thus enforcing safety. We verify our developed results on the marine vehicle based  on the Artemis Technologies Limited (ATL)\footnote{https://www.artemistechnologies.co.uk/ef-12-workboat/}  Simulink model on a host of benchmark Imazu scenarios. 

Notations and preliminaries are in Section 2. Section 3 discusses the control problem formulation. The main results describing the proposed guidance law in waypoint reaching mode  in cascade with the predefined-time heading autopilot are provided in Section 4. The analysis of static obstacles avoidance is presented in Section 5.  Further results are derived for dynamic and multiple dynamic obstacle avoidance in Section 6. A detailed description of the implementation of the proposed results on a host of scenarios, namely, Imazu problems \cite{sawada2021automatic} and comparison with the existing methods are provided in Section 7. Concluding remarks are given in Section 8.
\section{Preliminaries}
\begin{figure}[H]
	\centering
	\includegraphics[width=\textwidth]{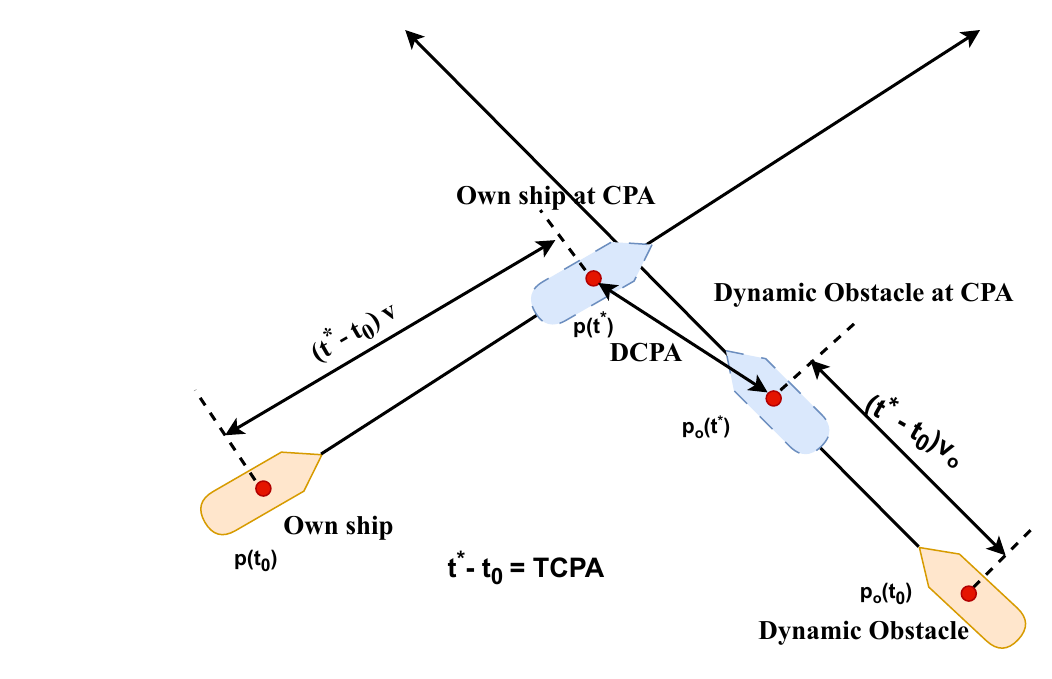} 	\caption{Illustration of CPA, DCPA and TCPA}
	\label{dc}
\end{figure}

 Let $\mathbb{R}$, $\mathbb{R}_{\geq p}$ and $\mathbb{R}_{>p}$ be the set of real numbers, real numbers greater than or equal to $p$ and real numbers greater than $p$ respectively. The complement of a set $S$ is $\bar{S}$. $\|\cdot\|$ denotes the Euclidean $2$-norm and $\|\cdot\|_{\infty}$ is the $\infty$-norm. If $x\in\mathbb{R},~x~\text{mod}~2\pi$ is modulo $2\pi$. We denote $\mathbb{S}^{1}$ as the set of real numbers of modulo $2\pi$. The notation $\arg\min$ denotes argument of the minimum. For instance, for $x\in\Omega$, $\arg\min\limits_{x}f(x)$ is the points $x$ in $\Omega$ for which $f(x)$ achieves its minimum value. For two sets $\cX$ and $\cS$, their Minkowski sum is $\mathcal{X}\oplus\mathcal{S}=\{ x+s, x\in\mathcal{X}, s\in\cS \}$. The convex hull between a set $\cS$ and a vector $x\in\R^2$ is $\conv{\cS,x}$. A ball induced by $l$-norm, centered at $y$ with radius $c$ is denoted as $\cB_l(y,c)=\{ p\in\mathbb{R}^2:~\|p-y\|_l\leq c\}$.
 We define $\text{atan2}:\mathbb{R}\times\mathbb{R}\to\mathbb{S}^{1}$ as
\begin{align*}
\text{atan2}(y,x):=~&\begin{cases}\text{tan}^{-1}\frac{y}{x},~~~~~~~~~~~~&\text{if}~x>0\\
		\text{tan}^{-1}\frac{y}{x}+\pi,~~~~~~~&\text{if}~x>0~\text{and}~y\geq 0\\
  \text{tan}^{-1}\frac{y}{x}-\pi,~~~~~~~&\text{if}~x<0~\text{and}~y<0\\
  \frac{\pi}{2},~~~~~~~~~~~~~~~~~~~&\text{if}~x=0~\text{and}~y>0\\
  -\frac{\pi}{2},~~~~~~~~~~~~~~~~~&\text{if}~x=0~\text{and}~y<0\\
  \text{undefined}~~~~~~~~~~~&\text{if}~x=0~\text{and}~y=0.
	\end{cases}
\end{align*}
We provide the illustration of CPA, DCPA, and time to closest point of approach (TCPA) in Figure \ref{dc} to calculate the risk of collision described in Section \ref{ri} during vessels encounters. The CPA, which can be evaluated geometrically, is the location where the distance between the two ships reaches minimum if they maintain their respective speeds and heading angles. DCPA shows the minimum distance between the own ship and the obstacle when the ship and the obstacle are at CPA, and TCPA is the time taken by the ship or obstacle to reach CPA. The mathematical expressions of DCPA and TCPA are in Section \ref{ri}.
\subsection{Marine Craft Dynamics}
\begin{figure}[H]\label{Artemis2}
	\begin{center}
		\includegraphics[width=\textwidth]{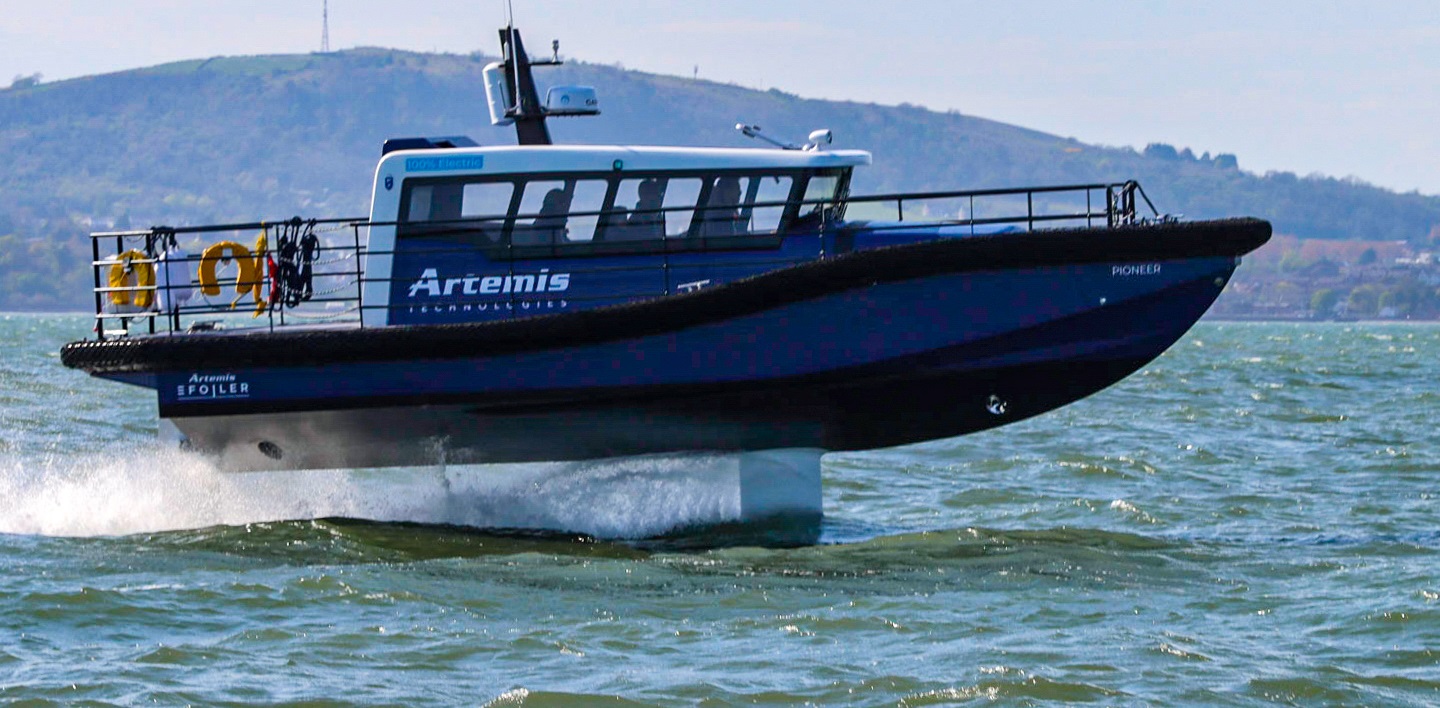}    
		\caption{The Artemis eFoiler EF-12 in the Belfast harbour.} 
		\label{fig:Artemis}
	\end{center}
\end{figure}

We consider the   marine vehicle dynamics 
\begin{align}\label{mar}
\begin{split}
\dot{x}(t)=&~v\cos(\psi(t)),\\
\dot{y}(t)=&~v\sin(\psi(t)),\\
\dot{\psi}(t)=&~-a\psi(t)+au(t),
\end{split}
\end{align}
where states $(x, y, \psi)$ take values in the set $\mathcal{X}\subset \mathbb{R}^{2}\times \mathbb{S}^{1}$ and  $u$ is the control that takes values in the set $\mathcal{U}\subset\mathbb{R}$. We denote $p(t)=[x(t),y(t)]^{\top}$ the position of the vehicle, $\psi(t)$ the heading angle, $v$ the constant velocity and $a$ a constant system parameter. This dynamic model of the ship is based on the ATL  closed-loop Simulink model provided for a constant velocity. A hydrofoiling marine vessel developed by ATL is shown in Figure \ref{fig:Artemis}.
We consider constraints on the heading rate  $|\dot{\psi}(t)|\leq c$, where $c>0$ is the permissible heading rate. From (\ref{mar}), we can induce constraints on the input  $|\dot{\psi}(t)|\leq a|\psi(t)|+a|u(t)|\leq a\pi+am\leq c$, or $u\in\mathcal{U}$, where
\begin{align}\label{cb}
    \mathcal{U}=\{u\in\mathbb{R}: |u|\leq m\},
\end{align}
 is the admissible control set and $m=\frac{c-a\pi}{a}$.  We consider  static predetermined waypoint target positions for the marine vehicle,  defined as $p_w=[x_w,y_w]^{\top}$.
\subsection{Obstacle Dynamics}
We consider  dynamics of the obstacles, or surrounding traffic, to be
\begin{align}\label{mo}
\begin{split}
\dot{x}_m(t)=&~v_o\cos(\psi_m),\\
\dot{y}_m(t)=&~v_o\sin(\psi_m),\\
\psi_m=&~\gamma,
\end{split}
\end{align}
where $(x_m,y_m,\psi_m)\in \mathbb{R}^{2}\times \mathbb{S}^{1}$ and $\gamma$ is a constant. We define $p_m(t)=[x_m(t), y_m(t)]^{\top}$ the position, 
$v_o$ is the constant velocity and $\psi_m$ is the constant heading angle of the moving obstacle. For multiple obstacles, we differentiate them with subscripts, e.g. $(x_{mi}, y_{mi}, \psi_{mi})$ for $i=1, \cdots, M$, where $M$ is the number of dynamic obstacles inside the detection range.
 We consider a sequence of waypoints that the ship should follow (given in our problem setup). For a static obstacle, $v_o=0$.

\subsection{Control Design Objectives}
 
 To safely reach the waypoint while avoiding obstacles with a safe distance, objectives (1)-(2) must hold simultaneously:

	1) \textbf{Waypoint (Goal) Reaching}: The marine vehicle must reach the waypoint $p^{\star}:=(x^{\star}, y^{\star})$   in finite time $T_F$, i.e.,
	$$p(t)\in \cB_2(p^{\star}, \delta), ~\text{for some}~ t\leq T_F,$$
	 and $\cB_2(p^{\star}, \delta)$ stands for the \emph{terminal set} (or COA)
  with $\delta>0 $ being a desirable accuracy.\\
	2) \textbf{Obstacle Avoidance}: The distance between the own ship $p(t)$ and any obstacle $p_m(t)$ (\ref{mo})  must be greater or equal to a safe distance $C_s>0$, that is,
	$$p(t)\notin \cB_{\infty}(p_m, C_s), $$
for all $t>0$,  where $\cB_{\infty}(p_m, C_s)$ is the unsafe set.
 We choose $C_s$ to be the smallest acceptable bound of the DCPA.
 
\subsection{Finite-time and Predefined-Time Stability}
We provide a brief note on finite-time and predefined-time stability, which we utilize to derive results of this work.
Consider a time-varying nonlinear system
 \begin{align}\label{sw}
 \dot{x}(t)=f(t,x(t), \sigma),~~~~~~~x(t_0)=x_0,
 \end{align}
 where $x(t)\in\mathbb{R}^n$ are the states, $\sigma\in\mathbb{R}^{p}$ denotes the vector of tunable constant system parameters and $f:\mathbb{R}_{\geq 0}\times\mathbb{R}^{n}\times\mathbb{R}^{p}\to\mathbb{R}^{n}$ is a nonlinear continuous function such that $f(t, 0, \sigma)=0$ for all $t\geq t_0$ and for all $\sigma\in\mathbb{R}^p$, that is, $x(t)=0$ is the equilibrium point of the system (\ref{sw}).
 \begin{definition}\cite{bhat2000finite}
     The origin of the system (\ref{sw}) is \emph{finite-time stable} if it is asymptotically stable for a fixed $\sigma\in\mathbb{R}^p$ and any solution trajectories of (\ref{sw}) converge to the origin in finite time, that is, $x(t, t_0, x_0)=0$ for all $t\geq t_0+T_f(t_0, x_0)$, where $T_f:\mathbb{R}_{\geq 0}\times \mathbb{R}^{n}\to \mathbb{R}_{\geq 0}$ is the time of convergence.
 \end{definition}
 \begin{definition}\cite{singh2022vector}
     The origin of the system (\ref{sw}) is \emph{predefined-time stable} if
          it is finite-time stable;
          it is possible to choose a  time of convergence duration $T_p=t_p-t_0>0$, which is independent of initial conditions and can be chosen \emph{a priori}; and
          the condition $T_t\leq T_p$ holds, where $T_t$ is the true convergence time duration in which the solution trajectories of (\ref{sw}) satisfies $x(T_t, t_0, x_0)=0$.  
 \end{definition}
 The following results are fundamental to analyze finite-time and predefined-time stability of nonlinear systems using Lyapunov functions.
  \begin{lemma}[Theorem 4.2, \cite{bhat2000finite}]\label{lem2}
  	Consider the system (\ref{sw}). Suppose there exists a positive definite continuously differentiable function $V:\mathbb{R}^{n}\to\mathbb{R}_{\geq 0}$ such that 
  	$$\dot{V}(x(t))\leq -\alpha V^{k}(x(t)), $$
  	for $\alpha>0$ and $0<k<1$. Then, the system (\ref{sw}) is finite-time stable about the origin and the time of convergence is given by $T_f=\frac{V(x(t_0))^{1-k}}{\alpha(1-k)}$.
  \end{lemma}
 \begin{lemma}\cite{singh2023finite}\label{lem1}
 	Consider the system (\ref{sw}). Suppose that there exists a positive definite continuously differentiable function $V:\mathbb{R}^{n}\to\mathbb{R}_{\geq 0}$ such that 
 	\begin{align*}
 	\dot{V}(x(t)) \leq
 	 -\frac{\eta (\text{e}^{V(x(t))}-1)}{\text{e}^{V(x(t))}(t_p-t)},~~~\text{for}~ t_0\leq t<t_p,
 	\end{align*} 
 	for  all $\eta\in\mathbb{R}_{>1}$, and $\dot{V}(x(t))=0~\text{for}~t\geq t_p$. Then the system (\ref{sw}) is predefined-time stable about the origin in time $t_p$.
 	\end{lemma}
  We consider the following predefined-time stable scalar dynamics (\ref{pre}) to establish convergence  and safety guarantees for our problem 
\begin{align}\label{pre}
	\dot{x}=\phi(t,x):=~&\begin{cases}-\frac{\eta (\text{e}^{x}-1)}{\text{e}^{x}(t_p-t)},~~~\text{for}~t_0<t<t_p,\\
		0,~~~~~~~~~~~~~~~~~~~~~~~~t\geq t_p,
	\end{cases}
\end{align}
where $x\in\mathbb{R}$ is the state,  $\eta\in\mathbb{R}_{>1}$ is a constant system parameter and $t_p=T_p+t_0$, where $T_p$ is the time duration prescribed \emph{a priori}.
This dynamics has a special feature, namely, $x=0$ and $\dot{x}=0$ for all $t\geq t_p$. Moreover, for a given value of $\eta$, the true convergence time of the dynamics  follows $T_t\leq T_p$, where $T_p$ is the preset of  settling time boundary. The detailed exposition can be found in \cite{singh2022vector}.
\begin{remark}
  It is worth noting that there are various options to select $\phi(t,x)$ in (\ref{pre}) for the predefined-time behaviour. One can also select $\phi(t,x)=-\frac{x}{T_p-t}$  \cite{song2017time}. However, our choice of $\phi (t, x)$ in (\ref{pre}) is preferable since it has a solution of nonlinear convergence and is defined for infinite time duration. 
\end{remark}

\begin{figure}[H]
	\centering
	\includegraphics[width=\textwidth]{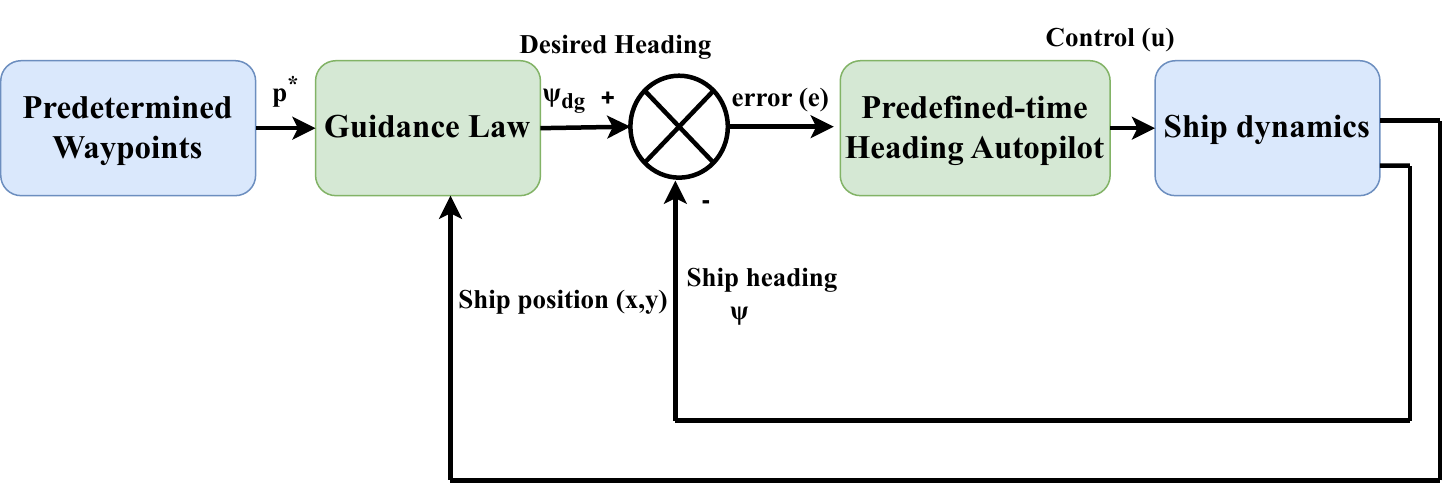} 	\caption{The proposed guidance and control system  for waypoint navigation and collision avoidance.}
	\label{fbl}
\end{figure}

The proposed  closed-loop guidance and control block diagram is shown in Figure \ref{fbl}.  Specifically, for a given sequence of waypoints (including virtual waypoints), we provide a guidance law to reach the goal   while avoiding static, dynamic (single and multiple) obstacles in its  path. We design a predefined-time heading controller such that the ship heading converges to the desired heading command provided by the guidance law, and as a result, safety is guaranteed. 

\subsection{Convention on the International Regulations (COLREGs)}
\begin{figure}[H]
	\centering
	\includegraphics[width=\textwidth]{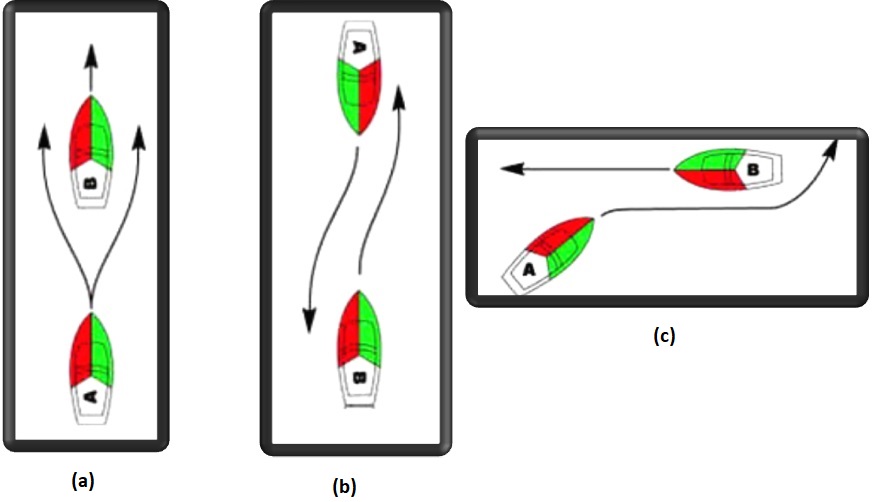} 	\caption{COLREGs rules on (a) Overtaking, (b) Head-on, and (c) Crossing (Source: \cite{cho2021intent})}
	\label{fblt}
\end{figure}

\begin{figure}[h!]
	\begin{center}
		\includegraphics[width=2in]{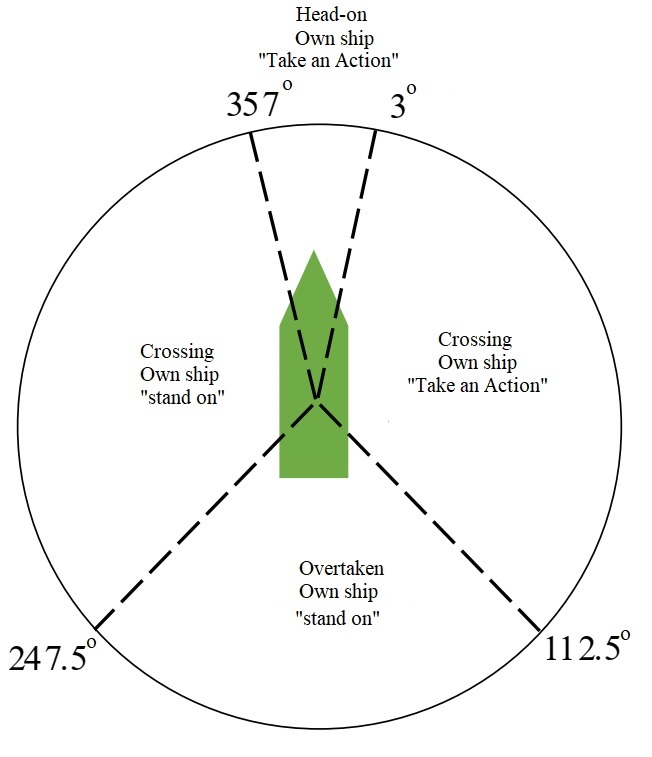}    
		\caption{Relative angles for collision risk assessment and COLREGs decision making.} 
		\label{fig:COLREGs}
	\end{center}
\end{figure}

During a journey, the ship must obey the Convention on the International Regulations (COLREGs) for preventing collisions at sea.  The COLREGs, are the `rules of the road' for maritime vehicles which are key to any sea-going vessel whether crewed or uncrewed. Rules 11-18  discuss situations when vessels are in sight of one another.  The three fundamental encounters are head-on (Rule 14), crossing (Rule 15) and overtaking scenarios (Rule 13) as shown in Figure \ref{fblt}. The action taken by the own ship is dependent  on its position relative to the target vessel according to Figure~\ref{fig:COLREGs}.

The own ship is either required to give way or stand-on depending on its relative position with the other vehicle. For example, in a crossing encounter, the own ship is required to give-way to another vessel which is on her own starboard side whilst the other ship is required to stand-on (Rule 15). For overtaking encounters (Rule 13), the ship that is being overtaken is required to stand-on, whereas the overtaking ship could pass from her port or starboard side with proper signaling.  The ship is considered as starboard-port symmetric. Generally, starboard maneuver is preferred over port-side. 
As previously mentioned, our control system has two different modes, namely, 
	(1) Waypoint reaching mode;
	(2) Obstacle avoidance mode.
The control system either follows the waypoint reaching guidance law discussed in Section \ref{nm} or obstacle avoidance guidance  described in Section \ref{tr}. 
In both these modes, we design the predefined-time heading autopilot controller such that the ship heading converges to the desired heading command (provided from guidance part) in pre-specified time $t_p$. The switching between the two modes occurs according to the criteria defined in Section \ref{sr}.

\section{Waypoint Reaching}\label{we}
In this section, the guidance law is provided for waypoint reaching. Further results are derived for the convergence of heading error in finite time $T_p$. Convergence of the ship position to the terminal set $\cB_2(p_w, \delta)$ is guaranteed in finite time $T_F>T_p$  using Lyapunov functions.
\subsection{Waypoint Reaching Guidance Law}\label{nm}
 
To reach the waypoint $p^{\star}:=(x^{\star}, y^{\star})$, we set the \textbf{desired guidance law (heading angle)} for (\ref{mar}) as
\begin{align}\label{ni}
\psi_{dg}=\text{atan2}\Big(y^{\star}-y,x^{\star}-x\Big).
\end{align}
We define the heading track error 
$$e(t)=\psi(t)-\psi_{dg}(t). $$
Consequently, the error dynamics is
\begin{align}\label{err}
\dot{e}(t)=-a\psi(t)+au(t)-\dot{\psi}_{dg}(t). 
\end{align}
The following result guarantees that the ship reaches the waypoint with the proposed control law $u$ while satisfying input constraints.
\begin{theorem}\label{the1}
Consider system (\ref{mar}), waypoint $p_w$ and the terminal set $\cB_2(p_w, \delta)$.\\
    (i) Consider the  control\footnote{We omit its dependence on time to ease notation.} $u(p, p_w)$,
    \begin{align}\label{con1}
u(p, p_w)=\begin{cases}
&\frac{1}{a}\dot{\psi}_{dg}+\psi-\frac{\eta(\text{e}^{(\psi-\psi_{dg})}-1)}{a\text{e}^{(\psi-\psi_{dg})}(t_p-t)},~ t_0\leq t<t_p,\\
&\psi,~~~~~~~~~~~~~~~~~~~~~~~~~~~~~~~~~\text{otherwise},
\end{cases}  
\end{align}
where $\psi_{dg}$ is given in (\ref{ni}).
	 Then, the heading track error dynamics (\ref{err}) is predefined-time stable.\\
(ii) In addition to (i),
$p(t)\in \cB_2(p_w, \delta)$ for some $t\leq T_F$, where $T_F=t_p+T_f$ and $$T_f=\frac{\sqrt{2}V_w(p(t_p))}{v}. $$ 
	(iii) Consider the input constraints $\mathcal{U}$ (\ref{cb}).  Then, $u(t)\in\mathcal{U}$ for all $t\geq t_0$ when $p(t_0)\in\bar{\cB}_2(p_w, \delta)\cup\bar{\cB}_{2}(p(t_0), vt_p)$, 

  and $\delta$ satisfies
  \begin{align}\label{no}
    \delta\geq \frac{2v}{a(m-\pi-\frac{\eta |(e^{\pi}-1)|}{a|(t_p-t_0)^{\eta}|})}.  
  \end{align} 
\end{theorem}
The proof can be found in the Appendix. Next, we present an example that illustrates Theorem \ref{the1}.\\
\begin{example}\label{ex1}
 Consider system (\ref{mar}) with controller  (\ref{con1}) satisfying (\ref{cb}) with $\psi_{dg}(p, p_w)$ in (\ref{ni}) and $(x(t_0), y(t_0), \psi(t_0))=(0,0,0)$, $p_w:=(x_w, y_w)=(0.0054,0.0048)$ nautical miles (nmi). We consider   $m=3$, $v=23.32$ knots, $\eta=3.5$, $a=1.67$ and set $t_p=1$ sec.  From the Figure \ref{f21t}, it is clear that error $e=\psi-\psi_{dg}$ is zero in $t_p=1$ sec and ship reaches the waypoint in  $1.148$ sec, thus validating (i)-(ii) of Theorem \ref{the1}. The ship starts  by considering the initial distance between the waypoint and the ship as
$\|p(t_0)-p_w\|=13.45\geq vt_p +\delta=12+0.3093=12.3093$ meters such that control is bounded $|u|\leq 3$, where $\delta$ is calculated from (\ref{ww}). The ship safely reaches the waypoint $p_w$ (when the distance $\|p_w-p(t)\|\leq \delta$) in Figure \ref{f21ts} validating (iii)  of Theorem \ref{the1}. \\

\begin{figure}[H]
	\centering
	\includegraphics[width=\textwidth]{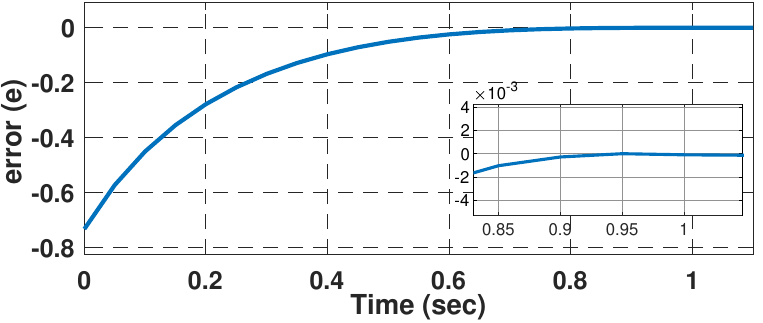} 	\caption{Example \ref{ex1}, convergence of heading error in predefined time $t_p=1$ sec}
	\label{f21t}
\end{figure} 
\begin{figure}[H]
	\centering
	\includegraphics[width=\textwidth]{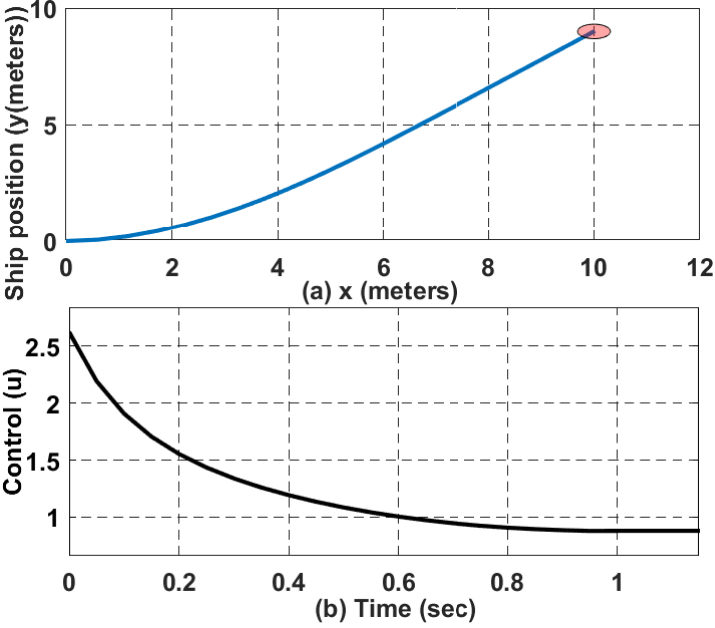} 	\caption{Example \ref{ex1}, ship position starting from ($0, 0$) to reach waypoint $(10, 9)$ meters terminal set (a red circle is $\cB_2(p_w, \delta)$)  with center at (10,9) and radius $\delta=0.3093$ meters) with bounded control $u$.}
	\label{f21ts}
\end{figure} 
\end{example}
\begin{remark}
  We note that the expression of $\dot{\psi}_{dg}$ (\ref{ds}) is analytic, and thus practicable for real-world implementation as there is no need to approximate the derivative.  
\end{remark}
\begin{remark}
  We note that $t_p=T_p+t_0$, is an independent parameter, which is
explicitly predefined in advance for a given value of a tuning parameter $\eta$. Theoretically, one can choose any
arbitrarily small value of $t_p$. However, practical systems impose restrictions on assuming arbitrarily small
values of $t_p$ due to inherent
limitations of the actuator constraints. 
\end{remark}
\begin{remark}
    Note that we could choose the asymptotic controller $u$ as in \cite{fossen2014uniform, wiig2019collision} (whereas we consider a point instead of a path in \cite{fossen2014uniform}), as a building block, i.e., 
$u=\frac{1}{a}\dot{\psi}_{dg}+\psi_{dg}, $ to make the heading track error dynamics (\ref{err})  asymptotically stable. The error follows $e(t)=e^{-at}e(0)$, $|\psi(t)-\psi_{dg}(t)|=e^{-at}|\psi(0)-\psi_{dg}(0)|$. 
 Then, after time ${T}_s\geq \frac{1}{a}\log_e(\frac{e(0)}{\epsilon_0})$, it holds $\psi=\psi_{dg}+\epsilon$, where $|\epsilon|\leq \epsilon_0$. Consequently,  (\ref{mar}) becomes
\begin{align}\label{see}
\begin{split}
    \dot{x}=&~v\cos(\psi_{dg}+\epsilon),\\
    \dot{y}=&~v\sin(\psi_{dg}+\epsilon).
    \end{split}
\end{align}
Choosing  Lyapunov function for (\ref{see}) as
$V_{aw}=\frac{1}{2}d_w^2$,
 we have
\begin{align*}
\dot{V}_{aw}=&-v(x_w-x)\cos(\psi_{dg}+\epsilon)-v(y_w-y)\sin(\psi_{dg}+\epsilon),\\
=&-v(x_w-x)(\cos(\psi_{dg})\cos(\epsilon)-\sin(\psi_{dg})\sin(\epsilon))\\&-v(y_w-y)(\sin(\psi_{dg})\cos(\epsilon)+\cos(\psi_{dg})\sin(\epsilon)),
\end{align*}
and by (\ref{er}), 
$\dot{V}_{aw}=-vd_w\cos\epsilon =-vd_w\cos(e^{-at}e(0))\implies \dot{V}_{aw}< ~0$, when ~$\cos\epsilon>0, ~\text{that is},~\text{when}~ -\pi<\epsilon<\pi.$\\

Thus, using an \emph{asymptotic} controller, one can only guarantee that the ship reaches the waypoint asymptotically, which is not sufficient for finite-time waypoint convergence.
\end{remark}

\section{Static Obstacle Avoidance}\label {nc}
In this section, we consider static obstacles. We discuss the switching rule between the waypoint reaching and collision avoidance modes. Next, we derive results that ensure that the vessel safely reaches the waypoint terminal set while maintaining a safe distance from the obstacle. Further, to avoid too large heading evasive maneuver in obstacle avoidance mode, we discuss another switching strategy between the two modes using risk index functions. We also discuss results for multiple static obstacles.

\subsection{Single Static Obstacle}\label{tr}

\begin{figure}[H]
	\centering
	\includegraphics[width=\textwidth]{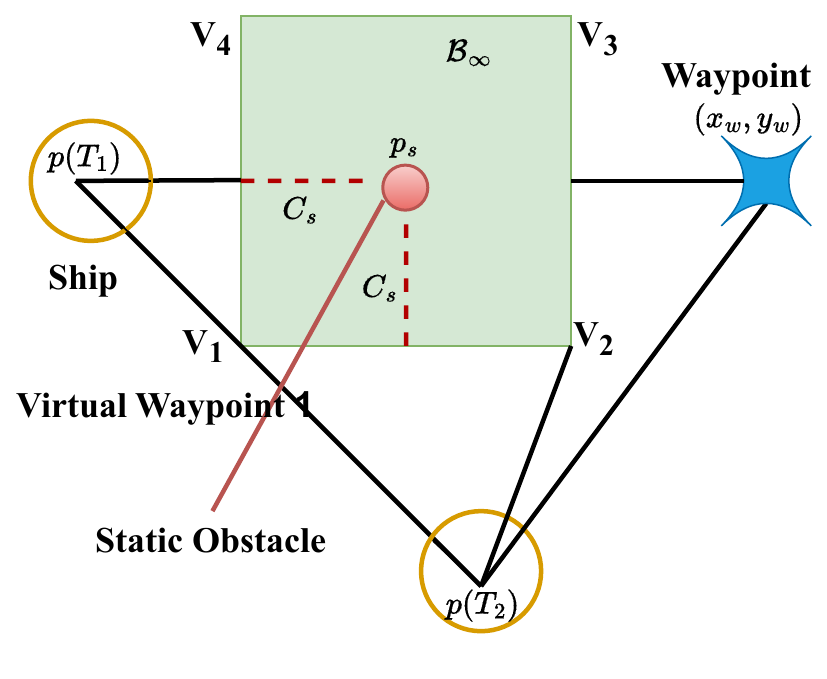} 
 \caption{Illustration of guidance law switching when one static obstacle is present}
	\label{f21}
\end{figure}

	 We consider a static obstacle  with the position $p_s:=(x_s, y_s)$, depicted in Figure \ref{f21} with a red circle enclosed in the set $\cB_{\infty}(p_s, C_s)$.
   We construct $\cB_{\infty}(p_s, C_s)$ by the given minimum allowable distance around the obstacle,  $C_s$, which corresponds to the smallest acceptable bound on DCPA. The ship can either take port-side or starboard maneuvers. There are four vertices of $\cB_{\infty}(p_s, C_s)$, namely, $V_i=[x_{V_i}, y_{V_i}]^{\top}$, $i=1, \cdots, 4$, out of which $V_1, V_2$ vertices can be used for port-side maneuvers and $V_3, V_4$ for starboard maneuvers. For port-side maneuvers, the relative angle $\varrho_{V_i, p}$ between the vertices $V_i$ and ship $p$ is positive and given by
  $$\varrho_{V_i, p}=\text{atan2}\Big((y_{V_{i}}-y(t)),(x_{V_{i}}-x(t))\Big)-\psi(t).$$ For starboard maneuvers, this angle becomes negative. 
 We also generate a virtual waypoint $V_1$. This virtual waypoint  is the vertex from which this relative angle $\varrho_{V_i, p}$ is largest for starboard or port-side maneuvers.  The set $\cB_2(p(t_0), vt_p)$ 
 denotes the \emph{transient circle} (yellow color circle) around the ship,
 and $d_{s}(t)=\|p_s-p(t)\|$ is the distance between the obstacle $(x_s,y_s)$ and the ship $(x,y)$. 
  We note that there always exists an affine transformation (change of variables)  that transforms the waypoint LOS alignment to the 
 $x$-axis, thus without loss of generality, $(x_s, y_s)=(x_s,0)$, $(x_{V_{1}}, y_{V_{1}})$ is calculated as $x_{V_{1}}=x_o-C_s,~ y_{V_{1}}=-C_s$, and any scenario can be brought to the configuration in Figure \ref{f21}.


\subsection{Switching Rule}\label{sr}
The switching modes for system (\ref{mar}) are depicted in Figure \ref{f21m}. This system has two modes: waypoint reaching mode (S1); and obstacle avoidance mode with two submodes S2, S3.\\
The control system is in mode S2 if the following conditions hold:
\begin{enumerate}
    \item [(1a)] there is intersection of $\cB_{\infty}(p_s, C_s)$ with the waypoint LOS set. The intersection condition is provided mathematically as follows. 
     We denote the ball around the ship position with a radius $v t_p$ as $\cB_2(p(t_0), vt_p)=\{p: \| p-p(t_0)  \| \leq v t_p  \}$. By Theorem \ref{the1}, we guarantee that the orientation of the ship is aligned with the LOS within at most $t_p$ seconds.  From the time $t=t_0$ when the  collision avoidance mode is activated, we can bound the trajectory of the ship $p(t)$ in the set $\cF(p(t_0))$, termed as waypoint LOS set, with
 \begin{align*}
\cF(p(t_0))=\conv{\cB_2(p(t_0), vt_p), p_w}.
 \end{align*}
    Consequently, the intersection condition mentioned above is stated as there being a time instant $t_0$ such that
\begin{equation} \label{cc1}
G_{11}:~~\cB_{\infty}(p_s, C_s)\cap\cF(p(t_0)) \neq \emptyset.
\end{equation}
\item [(1b)] The distance between the ship and the obstacle position $d_{s}=\|p-p_s\|$  satisfies $G_{12}$, where 
\begin{align}\label{df}
  G_{12}:~ d_{s}(t)\leq d_{safe}, 
\end{align}
and $d_{safe}= C_s+vt_p$.
\item [(1c)] The distance between the ship and the virtual waypoint $V_1$ follows $L_1$, where
\begin{align}\label{ed}
   L_1:~\|p(t)-V_1\|> \delta, 
\end{align}
 and $\delta$ is defined in (\ref{no}).
\item [(1d)] The virtual waypoint is ahead of the ship if $L_2$ holds
\begin{align}\label{gh}
\begin{split}
   L_{2}:~&-\frac{\pi}{2}<\text{atan2}\Big(y_{V_{1}}-y(t), x_{V_{1}}-x(t)\Big)-\psi(t)\\<~&\frac{\pi}{2}   
\end{split} 
\end{align}
\end{enumerate}
The control system is in S3 mode, if (\ref{cc1}), (\ref{df}), ($\bar{L}_1$ or $\bar{L}_2$) hold, where $$\bar{L}_2:~\frac{\pi}{2}<\text{atan2}\Big(y_{V_{1}}-y(t),x_{V_{1}}-x(t)\Big)-\psi(t)<\frac{3\pi}{2}. $$

The control system is in the waypoint reaching mode (S1), if the following conditions are satisfied:
\begin{enumerate}
\item [(2a)] There is intersection of the waypoint LOS set with $\cB_{\infty}(p_s, C_s)$, $G_{11}$ and the distance between the ship and the obstacle position $d_{s}=\|p-p_s\|$  satisfies $\bar{G}_{12}$. 
    \item [(2b)]There is a clear line-of-sight to the waypoint, that is, there is no intersection of the waypoint LOS set $\cF(p(t_0))$ with the set $\cB_{\infty}(p_s, C_s)$, i.e., $\bar{G}_{11}$.
    \end{enumerate}
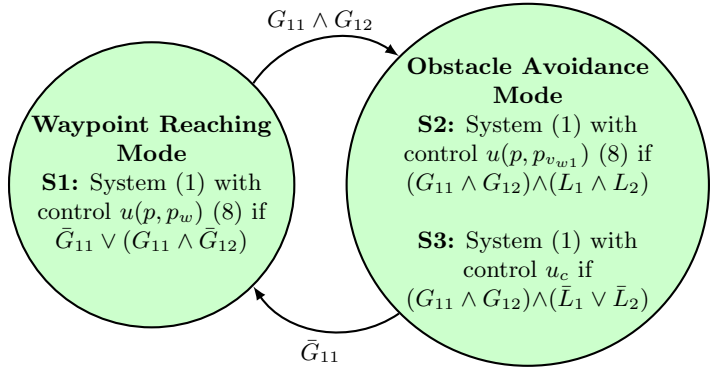
\begin{figure}[H]
\begin{tikzpicture}[
    node distance=5cm,
    on grid,
    thick,
    font=\small]
 
\node
[ 
    state,
    fill=green!20,
    align=center,
    inner sep=1pt
] (q_0) 
{
    \textbf{Waypoint Reaching}\\
    \textbf{Mode} \\ 
    \textbf{S1:} System (\ref{mar}) with\\  control $u(p, p_w)$ (\ref{con1}) if\\ $\bar{G}_{11} \lor ({G}_{11} \land \bar{G}_{12})$
};
 
\node
[
    state,
    fill=green!20,
    align=center,
    inner sep=1pt
] (q_1) [right=of q_0] 
{
    \textbf{Obstacle Avoidance}\\
    \textbf{Mode}\\
    \textbf{S2:} System (\ref{mar}) with\\ control $u(p, p_{v_{w1}})$ (\ref{con1}) if \\  $(G_{11} \land G_{12})$$\land (L_1\land L_2)$\\ \\
  \textbf{S3:}  System (\ref{mar}) with\\ control $u_c$ if\\
    $(G_{11} \land G_{12})$$\land (\bar{L}_1\lor \bar{L}_2)$

};

\path [-latex]
    (q_0) edge [bend left=45] 
        node [above] {$G_{11}\land G_{12}$ }(q_1)
    (q_1) edge [bend left=45]   
        node [below] {$\bar{G}_{11}$} (q_0);
\end{tikzpicture}
 \caption{Switching modes for the system (\ref{mar}) when one static obstacle is present.  The label on the edges describes the switching conditions.}
	\label{f21m}
\end{figure}

\noindent Thus, the overall switching controller $u$ for the system (\ref{mar}) is 
\begin{align}\label{cont}
    u=&~\begin{cases}
        u(p, p_w),~~~~~~~~\text{if}~~\text{S1,} \\
   u(p, p_{v_{w1}}),~~~~~~\text{if}~~\text{S2,}\\
        u_c,~~~~~~~~~~~~~~~\text{if}~~\text{S3.}
    \end{cases}
\end{align}
The controller $u(p, p_{v_{w1}})=u_c$ is constant until $\bar{L}_1\lor \bar{L}_2$ is satisfied.
\subsection{Safe Waypoint Reaching}
We show that the system (\ref{mar}) with the switching controller (\ref{cont})  safely reaches the waypoint terminal set in the presence of a static obstacle $p_s$, while maintaining a safe distance $C_s$ with the obstacle position. 
We pose the following  assumptions:

 \begin{assumption}\label{as:1}
   $d_s(t_0)>d_{safe}$. 
\end{assumption}
This assumption ensures that ship is safely initialized to start in the waypoint reaching mode before control is switched to the obstacle avoidance mode.
   \begin{assumption}\label{as:2}
   $d_{sw}(t_0)=\|p_s-p_w\|>d_{safe}$.
\end{assumption}
 Like Assumption 1, this assumption ensures that the ship is initialized such that it safely resumes to the waypoint reaching mode after avoiding obstacle.


\begin{theorem}\label{thee2}
Consider the system (\ref{mar}), waypoint $p_w:=(x_w, y_w)$, a static obstacle $p_s:=(x_s, y_s)$, set $\cB_{\infty}(p_s, C_s)$  and the terminal set $\cB_2(p_w,\delta)$.  Then, (i) the ship under the controller (\ref{cont}) reaches the waypoint terminal set $\cB_2(p_w,\delta)$ in finite time $T_F$; (ii) ship trajectory will not intersect with the unsafe set, that is,
  $$p(t)\notin \cB_{\infty}(p_s, C_s), ~~\forall ~ t\in[t_0, T_F].$$ 
\end{theorem}

\subsection{Switching Strategy based on Risk Assessment}\label{ri}

It is often desirable to avoid large and sudden heading evasive maneuvers in obstacle avoidance mode as it consumes large amount of energy and may cause discomfort to the passengers. For this reason, we offer another switching strategy based on risk assessment using fuzzy risk index such that ship can take an early and smooth action. As a result,  energy consumption is taken into account  while avoiding obstacles, unlike using a binary decision variable for risk assessment, which is a common method of choice in the literature \cite{campbell2012rule}.

The  DCPA  and TCPA are defined (see e.g. \cite{dastgerdi2023geometric}) as
\begin{align}\label{rr}
   {\text{DCPA}=\|R \|\sin (\alpha)}  ,~\text{TCPA}=\frac{\|R\|}{\|V_r\|},
\end{align}
where $\alpha=\text{cos}^{-1}(\frac{V_r^{\top}R}{\|V_r\|\|R\|}), V_r=[\dot{x}-\dot{x}_m, \dot{y}-\dot{y}_m]^{\top},~R=[x-x_m, y-y_m]^{\top}$. We perform switching between the modes using the risk index (RI) function  defined by
 \begin{equation}
 \text{RI}=\frac{1}{3}(F(\text{DCPA})+F(\text{TCPA})+F(d_s)), 
 \end{equation}
 where $F(\cdot):\mathbb{R}\to[0,1]$ is the fuzzy membership function (\cite{sarhadi2022}), 
 \begin{align}\label{eq: membership Function}
 \fontsize{7.7}{8.8}\selectfont
 F(z)~=
 \begin{cases}
 1, ~ & z\in [0,\beta_1],\\
 \\
 1-2\Big(\frac{z-\beta_1}{\beta_2-\beta_1}\Big)^{2},~& z\in  \Big(\beta_1, \frac{\beta_1+\beta_2}{2}\Big],\\
 \\
 2\Big(\frac{z-\beta_2}{\beta_2-\beta_1}\Big)^{2}, ~ & z\in  \Big(\frac{\beta_1+\beta_2}{2},\beta_2\Big],\\
 \\
 0, ~ &  z \in (\beta_2,\infty).\\
 \end{cases}
 \end{align}
 In  (\ref{eq: membership Function}), $z\in\mathbb{R}$ denotes DCPA,  TCPA or distance between the ship and the obstacle ($d_s=\|p(t)-p_s\|$)  and $\beta_1, \beta_2$ are the tuning parameters for these metrics. 
 The control system is in the obstacle avoidance mode, if 
 condition (\ref{cc1}) and $G_{22}$ hold,
 \begin{align}\label{ss}
    G_{22}:~~ RI\geq K,
 \end{align}
 where $K\in[0,1]$ is a constant value.
The control system resumes the waypoint reaching mode similar to Section \ref{sr}.
Consequently, the overall switching controller $u$ for the system (\ref{mar}) is (\ref{cont}) when $G_{12}$ is replaced by $G_{22}$.

\begin{remark}
In this switching strategy, it is worth noting that the control system  enters the obstacle avoidance mode when $G_{22}$ is satisfied, at which the distance of the ship from the obstacle is $d_s(t)\leq d_{safe}$,  where $\beta_1<d_{safe}<\beta_2$, and $\beta_1=C_s+vt_p$. In general, any condition $G_{22}$, where $d_{safe}$ satisfies $\beta_1<d_{safe}<\beta_2$ is sufficient so that the statements of Theorem \ref{thee2} would hold by replacing $G_{12}$ with $G_{22}$.  
\end{remark}

 \begin{example}\label{ex2}
    Consider system (\ref{mar}) with control (\ref{cont}) with $(x(t_0), y(t_0), \psi(t_0))=(0,0,0)$ and $p_w:=(x_w, y_w)=(500,0)$ meters. We consider values of constants as  $m=18, v=12 ~\text{m/sec}, \eta=3.5, t_p=1~ \text{sec}, a=1.67 $ and static obstacle with position $(200,0)$ meters with $C_s=50 $ meters. The new intermediate virtual waypoint in obstacle avoidance mode is calculated as $(x_{V_{1}},y_{V_{1}})=(150, -50)$ meters. The ship enters the obstacle avoidance mode from the waypoint reaching mode as the conditions (\ref{cc1})-(\ref{df}) are satisfied with the minimum or worst-case distance between the obstacle and ship position $ d_{s}(t)\leq d_{safe}=vt_p+C_s=12+50=62$ meters when $u$ is (\ref{cont}), and when $G_{12}$ is replaced by $G_{22}$ (\ref{ss}) in (\ref{cont}), with
 $RI_1=0.35, RI_2=0.5$. \\
 
 From the Figure \ref{fst}, it is clear that $d_s(T_1+t_p)\geq C_s$ for (\ref{cont}) (binary decision -- bd) and the modified risk assessment based switching strategies and the ship remains at a distance greater than or equal to $C_s$ until it again enters the waypoint reaching mode. With the condition $\bar{G}_{11}$, it resumes to the waypoint reaching mode and hence it is evident that ship trajectories reach the waypoint terminal set and never come inside the  unsafe  set $\cB_{\infty}(p_s, C_s)$  around the obstacle that validates Theorem \ref{thee2}. The control input and the heading angle of the ship are shown in Figure \ref{fstr} where $T_1=11.5s, t_p=1s, T_2=18s$  in  the obstacle avoidance and waypoint reaching mode respectively. Thus,  $\psi=\psi_{dg}$ in just $1$ second in obstacle avoidance and   waypoint reaching mode. 
 \begin{figure}[H]
	\centering
	\includegraphics[width=\textwidth]{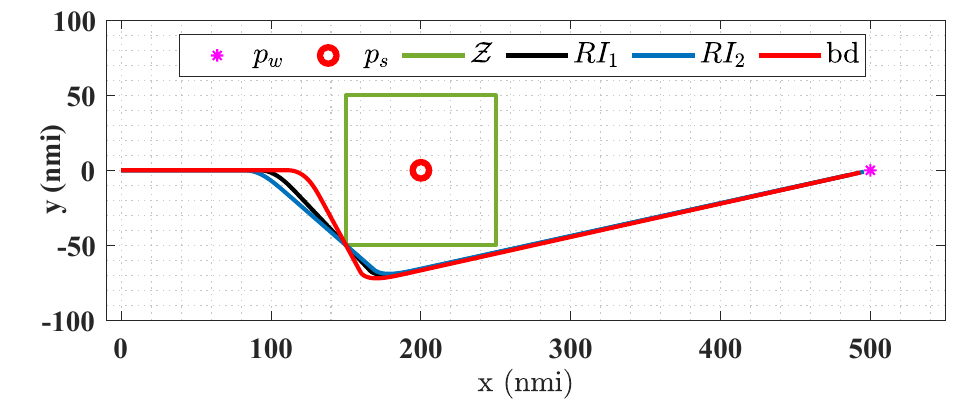} 	\caption{Example \ref{ex2}, ship maneuvering in the presence of static obstacle at $(200,0)$ meters at a safe distance $C_s$ and reaches the waypoint at $(500,0)$ meters, where $RI_1=0.35, RI_2=0.5$.}
	\label{fst}
\end{figure} 
\begin{figure}[H]
	\centering
	\includegraphics[width=\textwidth]{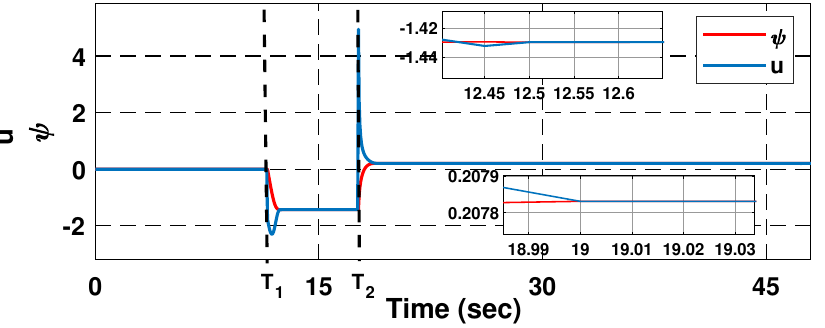} 	\caption{Example \ref{ex2}, ship heading and control input with $T_1=11.5$s,  $T_2=18$s,  and $t_p=1$s for binary decision.}
	\label{fstr}
\end{figure} 
 \end{example}
  \subsection{Multiple Static Obstacles}
  The developed results can be extended to multiple static obstacles. A detailed formal analysis of all possible scenarios is the object of our future work. We discuss a rule such that ship navigates waypoint while avoiding safely multiple static obstacles.
   If multiple static obstacles $p_{si}=[x_{si} \ \ y_{si}]^{\top},~i=1, \cdots, M$ are present, and the distance between any two static obstacles with the positions $p_{si}=[x_{si} \ y_{si}]^{\top},p_{sj}=[x_{sj} \ y_{sj}]^{\top},~i=1, \cdots, M, j=1, \cdots, M, i\neq j $, $d_{s_{ij}}=\|p_{si}-p_{sj}\|<2d_{safe}, ~\forall~ i,j$, where $d_{safe}=C_s+vt_p$. In such cases, obstacles are regarded as clustered and the set $\mathcal{Z}$ with center $p_{sc}=[x_{sc}, y_{sc}]^{\top}$ is constructed as follows
   \begin{align*}
       \mathcal{Z}=\{(x,y)|~|x- x_{sc}|\leq L~|y-y_{sc}|\leq B\},
   \end{align*}
   with $L=\frac{1}{2}(\max\limits_{i=1, \cdots, M} x_{si}-\min\limits_{i=1, \cdots, M} x_{si}+2C_s), ~B=\frac{1}{2}(\max\limits_{i=1, \cdots, M} y_{si}-\min\limits_{i=1, \cdots, M} y_{si}+2C_s)$.
   \begin{figure}[H]
	\centering
	\includegraphics[width=6cm]{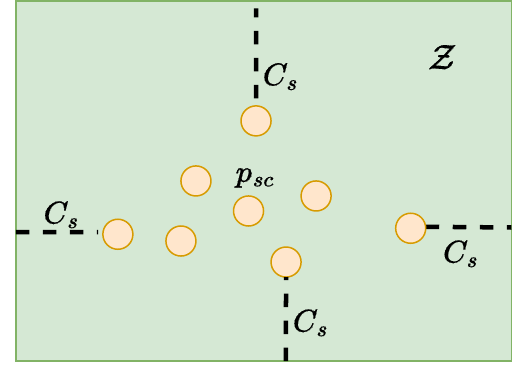} 
 \caption{Construction of unsafe set $\mathcal{Z}$ (green colored box) for clustered static obstacles (yellow colored circles) where $\|p_{si}-p_{sj}\|<2d_{safe}$.}
	\label{multi_static}
\end{figure} 


 Theorem \ref{thee2} results hold for multiple static obstacles with the only difference that the set $\mathcal{Z}$ is a rectangle and not a square as in the case of a single static obstacle (e.g., Figure \ref{f21}).
 \begin{remark}
     We note in the definition of the unsafe set $\cB_{\infty}(p_s, C_s)$, the norm used is infinity norm.
     In the case where multiple obstacles are present, the use of convex hull of the vertices to reduce the size of the unsafe set is an option for future work.
 \end{remark}
\section{Dynamic Obstacle Avoidance}\label{mg}
\begin{figure}[H]
	\centering
	\includegraphics[width=9cm]{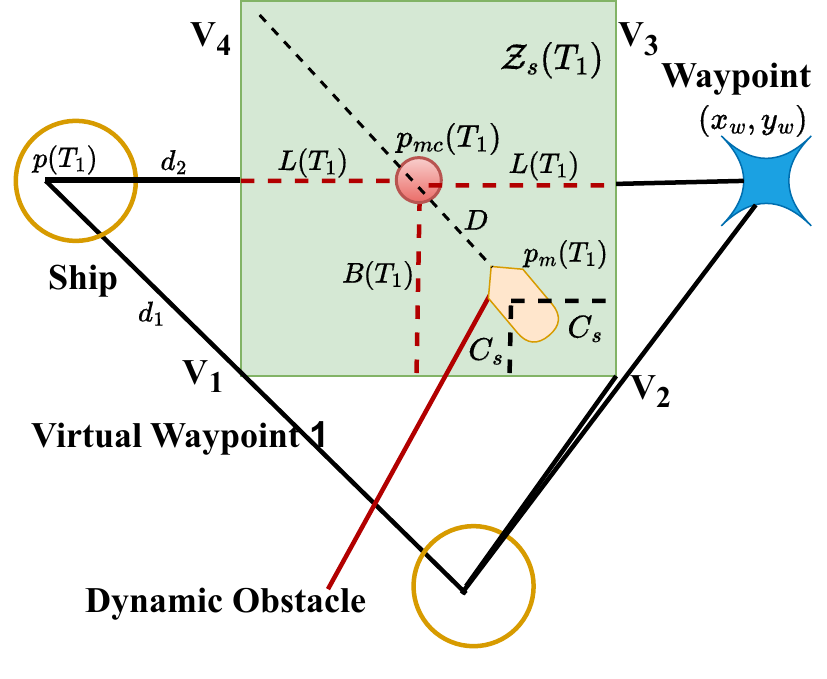} 	\caption{Illustration of guidance law switching when one dynamic obstacle is present (in the fourth quadrant)}
	\label{f213}
\end{figure}

 In this section, we consider a dynamic obstacle  with the position $p_m(t)=[x_m(t), y_m(t)]^{\top}$, represented by system (\ref{mo}), depicted in Figure \ref{f213} at time $T_1$. It is enclosed in the  set $\mathcal{Z}_s(T_1)$, where  
 \begin{align}\label{zz}
 \begin{split}
    \mathcal{Z}_s(t)= \left\{(x,y)~|~|x-x_{mc}(t)|\leq L(t),|y-y_{mc}(t)|\leq B(t)\right\},
 \end{split}   
 \end{align}
and $p_{mc}(T_1)=[x_{mc}(T_1) \  y_{mc}(T_1)]^{\top}$ is the CPA at $T_1$, $B(T_1)=|y_{m}(T_1)-y_{mc}(T_1)|+C_s$, $L(T_1)=|x_{m}(T_1)-x_{mc}(T_1)|+C_s$.  The coordinates of the closest point of approach (CPA) $p_{mc}(T_1) $  at $t=T_1$ is
\begin{align*}
    x_{mc}=&~ v_o\cos(\psi_m)T_c+x_m(T_1),\\ y_{mc}= &~v_o\sin(\psi_m)T_c+y_m(T_1),
\end{align*}
where $T_c$ is the TCPA as defined in (\ref{rr}).
The minimum allowable distance around the dynamic obstacle $p_m(T_1)$ is $C_s$. The distance between the CPA $p_{mc}(T_1)$ corresponding to $t=T_1$ and the ship $p(t)$ is $d_{s}(t)=\|p(t)-p_{mc}(T_1)\|$. 
The four vertices of $\mathcal{Z}_s(T_1)$ are $V_i(T_1)=[x_{V_{i}}(T_1) \  y_{V_{i}}(T_1)]^{\top}$ $i=1,...,4$.  Thus, we consider as the virtual waypoint the vertex $V_1(T_1)$ as described in Section \ref{tr}. For simplicity of exposition and without loss of gnenerality, we set the $x$-axis to the initial LOS alignment leading to  $(x_{mc}, y_{mc})=(x_{mc},0)$,  $x_{V_{1}}(T_1)=x_{mc}-|x_{m}(T_1)-x_{mc}|-C_s, ~y_{V_{1}}(T_1)=-|y_{m}(T_1)|-C_s$. 
We note that the ship $p(t)$ and obstacle $p_m(t)$ reach the CPA $p_{mc}(T_1)$ in time TCPA. Thus,
\begin{align}\label{fv}
\begin{split}
    &D=~\sqrt{(x_m(T_1)-x_{mc}(T_1))^2+y_m^2(T_1)},\\
    &d_1=(\text{TCPA}-t_p)v,\\
   & d_2=~\sqrt{\Big(\frac{D}{v_o}-t_p\Big)^2v^2-(y_m(T_1)+C_s)^2},
    \end{split}
\end{align} 
where $D,d_1,d_2$ are shown in Figure \ref{f213}. We consider  S2 and S3 modes according to the conditions described in Section \ref{sr}, 
 by setting $\cB_{\infty}(p_s, C_s)=\mathcal{Z}_s(T_1)$, $p_s=p_{mc}(T_1)$, $V_1=V_1(T_1)$, $d_{safe}=d_2+L(T_1)+vt_p, ~vt_p\leq v_rt_p, ~v_r=v-v_o$, and $T_1$ is the time instant when the conditions (\ref{cc1})-(\ref{df}) are triggered. 
 
 After reaching the virtual waypoint $V_1(T_1)$, the ship can either trigger one additional obstacle avoidance maneuver, or resume to the waypoint reaching mode if it can be guaranteed there is no intersection of its predicted trajectory with the dynamic obstacle. Technically, this is equivalent to guaranteeing there is an empty intersection between the trajectory of the obstacle, inflated by the safe required distance $C_s$, with the
possible trajectory of the ship for the whole duration of the waypoint reaching maneuver. In particular, we define the unsafe set $\mathcal{F}_0$ as follows
\begin{align}
\mathcal{F}_0(p_m(t_0), & T) =\{ p: \eqref{mo} \text{ holds}, \exists~ t \in[t_0,t_0+T]: \nonumber\\
 & p_m=[x_m(t) \ \ y_m(t)]^\top \}\oplus \cB_{\infty}(0,C_s). \label{eq_obsreach}
\end{align}
 We denote the ball around the ship position with a radius $v t_p$ as $\cB_2(p(t_0), vt_p)$. By Theorem \ref{the1}, we guarantee that the orientation of the ship is aligned with the LOS within at most $t_p$ seconds.  From the time $t=t_0$ when the  waypoint reaching mode is activated, we can bound the trajectory of the ship $p(t)$ in the set $\cF(p(t_0))$, with
 \begin{align}
\cF(p(t_0))=\conv{\cB_2(p(t),vt_p), p_w}.
 \end{align} \label{eq_shipreach}
Consequently, the empty intersection condition mentioned above is stated as there being a time instant $t_0$ so that
\begin{equation} \label{eq_emptyins}
G_{23}:~~\cF_0(p_m(t_0),T)\cap\cF(p(t_0)) =\emptyset,
\end{equation}
where $T=t_p+\frac{v\sin(\psi_{los})}{C_s+vt_p}, ~\psi_{los}=\text{atan2}(y_w-y(t_0), x_w-x(t_0))$, $t_0$ being the time instant when the waypoint resuming condition ${G}_{23}$ is triggered. 
 Thus, the closed-loop system resumes to the waypoint reaching mode S1 if \eqref{eq_emptyins} is satisfied.
Consequently, the overall switching controller $u$ for the system (\ref{mar}) is (\ref{cont}) by replacing $\bar{G}_{11}$ with $G_{23}$ \eqref{eq_emptyins}. 
\begin{figure}[H]
	\centering
	\includegraphics[width=\textwidth]{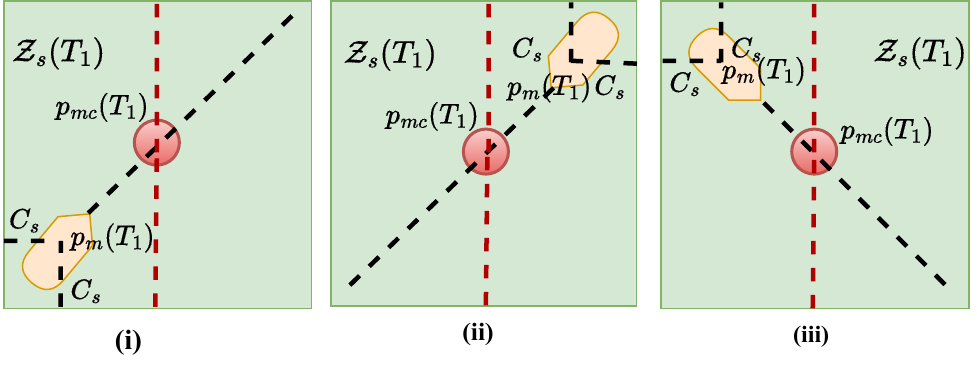} 	\caption{Possible motions of the dynamic obstacle $p_m$ (i) third quadrant, (ii) first quadrant, (iii) second quadrant.}
	\label{f213r}
\end{figure} 
We provide the following result for a dynamic obstacle whose possible motions are shown in Figure \ref{f213} or \ref{f213r} (i)-(iii). 
\begin{theorem}\label{thee3}
Consider system (\ref{mar}) at time $t_0$, waypoint $p_w$, a dynamic obstacle $p_m$  (\ref{mo}),  the terminal set $\cB_2(p_w, \delta)$, sets  $\mathcal{Z}_s(T_1)$ (\ref{zz}) and $\cB_{\infty}(p_m, C_s)$.  
Suppose Assumptions 1, 2 hold. Suppose that after a possible obstacle avoidance maneuver, the ship and obstacle trajectories are not parallel, or, if they are parallel the following condition holds
\begin{align}\label{dfR}
    \frac{v}{v_o}\neq~\frac{S+(C_s+v_rt_p)}{S},
\end{align}
with $S=\|p_w-p_{m}(t_w)\|$, and $t_w\geq T_1+T_c$ denotes the time when $p,p_m$ initiate parallel trajectories. Then: 

\noindent  (i) the ship under the controller (\ref{cont}) reaches  $\cB_2(p_w, \delta)$ in finite time $T_F$.

\noindent (ii) the ship is always safe, i.e., it will not intersect with the unsafe set so that
 $p(t)\notin \cB_{\infty}(p_m, C_s), ~\forall ~ t\in[t_0, T_F]$.
\end{theorem}

\begin{remark}
It is worth noting that condition (\ref{dfR}) is not satisfied for a single configuration of the dynamic obstacle and ship velocities. If this is the case, it suffices to add a small perturbation in ship's velocity to enforce it. 
\end{remark}

\begin{remark}
We note that the conditions needed to be verified for the dynamic obstacle case are similar to the static obstacle, with additional checks requiring computation of the sets \eqref{eq_shipreach}, \eqref{eq_obsreach}, that can be straightforwardly calculated.
\end{remark}

\subsection{Multiple Dynamic Obstacles }
In this subsection, we consider the most general case where a multitude of moving obstacles can be present that simultaneously affect the guidance algorithm as they trigger an obstacle avoidance maneuver. We present a procedure to address this setting by modifying the above results. We mention that the algorithm will guarantee obstacle avoidance under a key assumption, however extensive simulations, including the challenging Imazu scenarios (Cases 5-22) illustrated in the subsequent section VII show that the suggested algorithm is successful in more general settings. It is the object of our future research to formally prove the validity of the algorithm to this most general case. 

We consider $M$ dynamic obstacles, where each $i$-th obstacle can be represented by the system (\ref{mo}), where the position is $p_{mi}(t)=[x_{mi}(t) \ \ y_{mi}(t)]^{\top},~i=1, \cdots, M$, 
and the constant heading angle is
$\psi_{mi}$. For simplicity, we consider the same velocity $v_o$ for all obstacles. The minimum allowable distance around each dynamic obstacle $p_{mi}$ is $C_s$, which corresponds to the smallest acceptable bound on DCPA.  We  construct the set $\mathcal{Z}_d(t)$ that includes the dynamic obstacles as follows. 
\begin{enumerate}
    \item [(1)] Calculate $D_{mi}(t)=\|p(t)-p_{mi}(t)\|$. If $D_{mi}(t)<D_{sf}=C_s+vt_p, ~ vt_p=\|p(t_p)-p(t_0)\|\leq v_rt_p$, then $i$-th obstacle $p_{mi}(t)$ is inside the set $\mathcal{Z}_d(t)$; 
    \item [(2)] Calculate $D_{mij}(t)=\|p_{mi}(t)-p_{mj}(t)\|, i\neq j$. If $D_{mij}(t)\leq 2D_{sf}$, then $j$-th obstacle $p_{mj}(t)$ is inside the set $\mathcal{Z}_d(t)$;
     \item [(3)] Calculate $(TCPA_k, CPA_k)$ for all $k=1, \cdots, M$. If $\|p(t)-CPA_k\|<D_{sf}$, then $k$-th obstacle $p_{mk}(t)$ is in the set $\mathcal{Z}_d(t)$;
      \item [(4)] If $c=\arg\min\limits_{k=1, \cdots, M}\|p(t)-CPA_k\|$ corresponding to $(CPA_c, TCPA_c)$, then calculate the future position of all obstacles at $t=TCPA_c$, i.e.,  $p_{ml}=[x_{ml}(TCPA_c) \ \ y_{ml}(TCPA_c)]^\top$, $l=1, \cdots, M$ obstacles. If $\|p(t)-p_{ml}\|<D_{sf}$, then  $l$-th obstacle $p_{ml}(t)$ is inside $\mathcal{Z}_d(t)$.
\end{enumerate}
After these four steps, if there are $Q$ number of obstacles in the set $\mathcal{Z}_d(t)$, then the set $\mathcal{Z}_{d}(t)$ is constructed  as follows:
\begin{align*}
 \begin{split}
    \mathcal{Z}_d(t)=&\{(x,y)|~|x-x_{mc}(t)|\leq L(t),|y-y_{mc}(t)|\leq B(t)\}. 
 \end{split}   
 \end{align*}
where $p_{mc}(t)=[x_{mc}(t) \ \ y_{mc}(t)]^\top$ is the $CPA$ of the obstacle  which has minimum distance to the ship termed as the \emph{most risky obstacle}, with $B(t)=\text{max}\{|y_{mq}(t)|\}+C_s$ and $L(t)=\text{max}\{|x_{mq}(t)-x_{mc}(t)|\}+C_s,~q=1, \cdots, Q$. 

\begin{figure}[H]
	\centering
	\includegraphics[width=\textwidth]{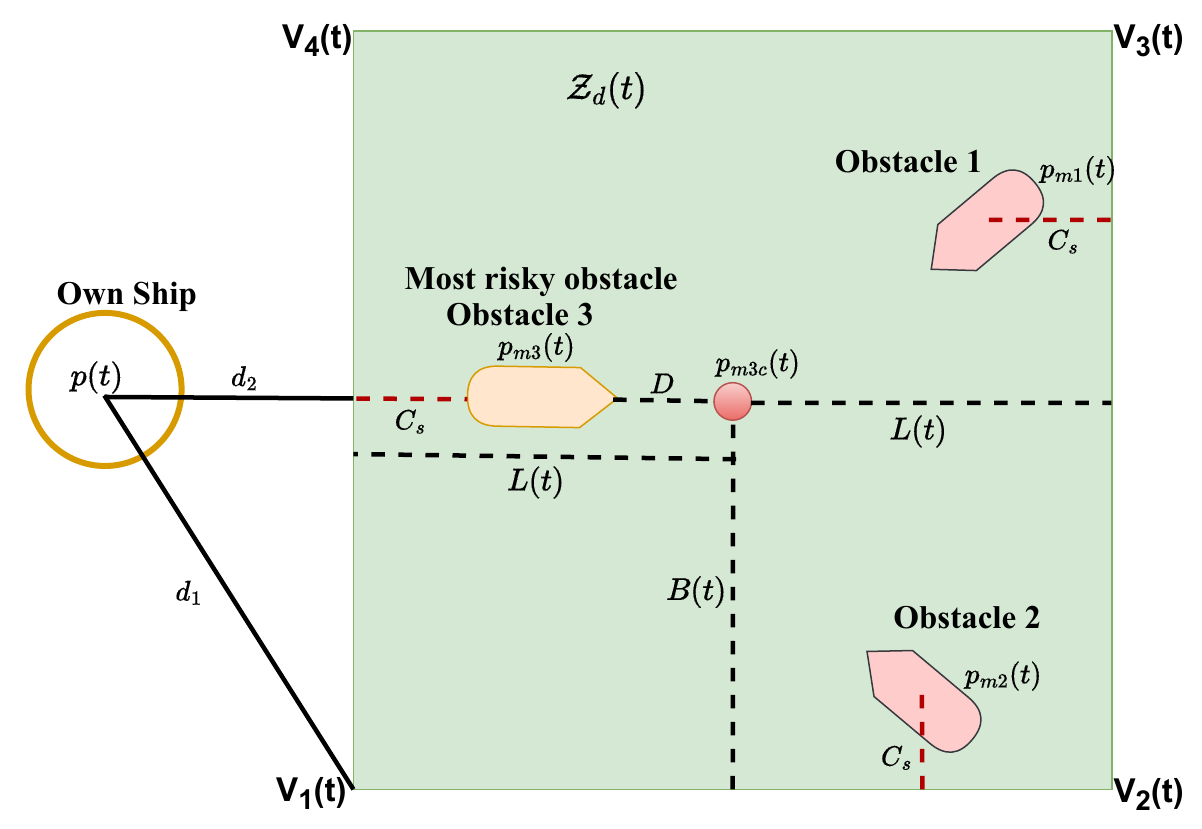} 	\caption{Illustration of switching of guidance law when multiple obstacles are present}
	\label{f218}
\end{figure}

To give an example, in Figure \ref{f218}, we consider $Q=3$, and obstacle 3's  CPA has minimum distance to the ship at time instant $T_{11}$, thus, $p_{mc}=p_{m3c}(T_{11})=[x_{m3c}T_{11}) \ \ y_{m3c}(T_{11})]^\top$, $B(T_{11})=|y_{m2}(T_{11})|+C_s$, $L(T_{11})=|x_{m1}(T_{11})-x_{m3c}(T_{11})|+C_s$, and $\mathcal{Z}_d=\mathcal{Z}_d(T_{11})$. The distance between the CPA of the most risky obstacle $p_{m3c}(T_{11})$ corresponding to $t=T_{11}$ and the ship $p(t)$ is $d_{s}(t)=\|p(t)-p_{m3c}(T_{11})\|$. 
The four vertices of $\mathcal{Z}_d(T_{11})$ are $V_i(T_{11})=[x_{V_{i}}(T_{11}) \ \  y_{V_{i}}(T_{11})]^\top$, $i=1,...,4$.  Thus, we consider as the virtual waypoint the vertex $V_1(T_{11})$ as described in Section \ref{tr}. 
We note that ship $p(t)$ and obstacle $p_{m3}(t)$ reach the CPA $p_{m3c}$ in time $\text{TCPA}$. Thus, $D=~|x_{m3}(T_{11})-x_{m3c}(T_{11})|$,  $d_1=(\text{TCPA}-t_p)v,$ $ d_2=~\sqrt{\Big(\frac{D}{v_o}-t_p\Big)^2v^2-(y_{m2}(T_{11})+C_s)^2}$,
where $D,d_1,d_2$ are shown in Figure \ref{f218}.\footnote{As in Theorem \ref{the1}, we note  there always exists an affine transformation  that transforms the waypoint LOS alignment to the 
 $x$-axis. Thus without loss of generality, we set
 $(x_{m3c}, y_{m3c})=(x_{m3c},0)$,  $x_{V_{1}}(T_{11})=x_{m3c}-|x_{m1}(T_{11})-x_{m3c}|-C_s, ~y_{V_{2}}(T_{11})=-|y_{m2}(T_{11})|-C_s$. Consequently, any case can be brought to the configuration in Figure \ref{f218}.}
 
  We consider the control \eqref{cont}, with S2 and S3 modes according to the conditions described in Section \ref{sr}, 
 by setting $\cB_{\infty}(p_s, C_s)=\mathcal{Z}_d(T_{11})$, $p_s=p_{m3c}(t)$, $p_{v_{w1}}=V_1(T_{11})$, $d_{safe}=d_2+L(T_{11})+vt_p, ~vt_p\leq v_rt_p, ~v_r=v-v_o$, and $T_{11}$ is the time instant when the conditions (\ref{cc1})-(\ref{df}) are triggered and obstacle 3 is most risky obstacle and $\mathcal{Z}_d(T_{11})$ is calculated according to $p_{m3c}(T_{11})$. Thus, Theorem \ref{thee2} results hold for this case as well provided the obstacle 3 remains risky throughout the obstacle avoidance mode. This implies that $p_{m3c}$ remains the same (constant) until the ship reaches to the virtual waypoint $V_1(T_{11})$ of set $\mathcal{Z}_d(T_{11})$ in some time finite time $T_{F1}$ according to Theorem \ref{thee3}.
 After the time $T_{F1}$, the ship
will be able to either to return to the waypoint reaching mode or engage in additional obstacle avoidance manouvers as in Theorem 3, again under the assumption that the most risky obstacle remains the same during each obstacle avoidance maneuver.
\section{Result and Discussion}
\subsection{Simulation for Imazu problems}
We verify our results on the high-speed hydrofoiling ship model (\ref{mar}) derived by identifying the Artemis Technologies Limited (ATL) Simulink model, to a host of scenarios, namely Imazu problems \cite{sawada2021automatic}. 
Imazu problems include worst case scenarios of the ship encounters, of which cases 1-4 are single ship encounter scenario,  5-11 are two ships encounters, whilst 12-22 involve three ship encounters. \\

We consider the ship initial position as $[x(t_0) \ \ y(t_0)]^\top=[0 \ \ 0]^\top$, heading angle $\psi(t_0)=0$ degrees and velocity $v=25$ knots in all the considered scenarios. The position of the  waypoint is $p_w=[12.82 \ \ 0]^\top$ nautical miles (nmi) and velocity of all dynamic obstacles is $v_o=10$ knots. For risk assessment, we set the value of $K$ as 0.35 nmi, DCPA $\in [0.25, 0.5]$ nmi, TCPA $\in [2, 4]$ minutes, and obstacle distance: $\|D\|\in [0.08, 0.25]$ nmi. The initial position and heading angle ($p_o(t_0), \psi_m$) of the target ships (dynamic obstacles) for all 22 cases of Imazu problems are provided in Table \ref{tb1}. \\

\begin{figure}[h!]
\includegraphics[height=6.5in, width=\textwidth]{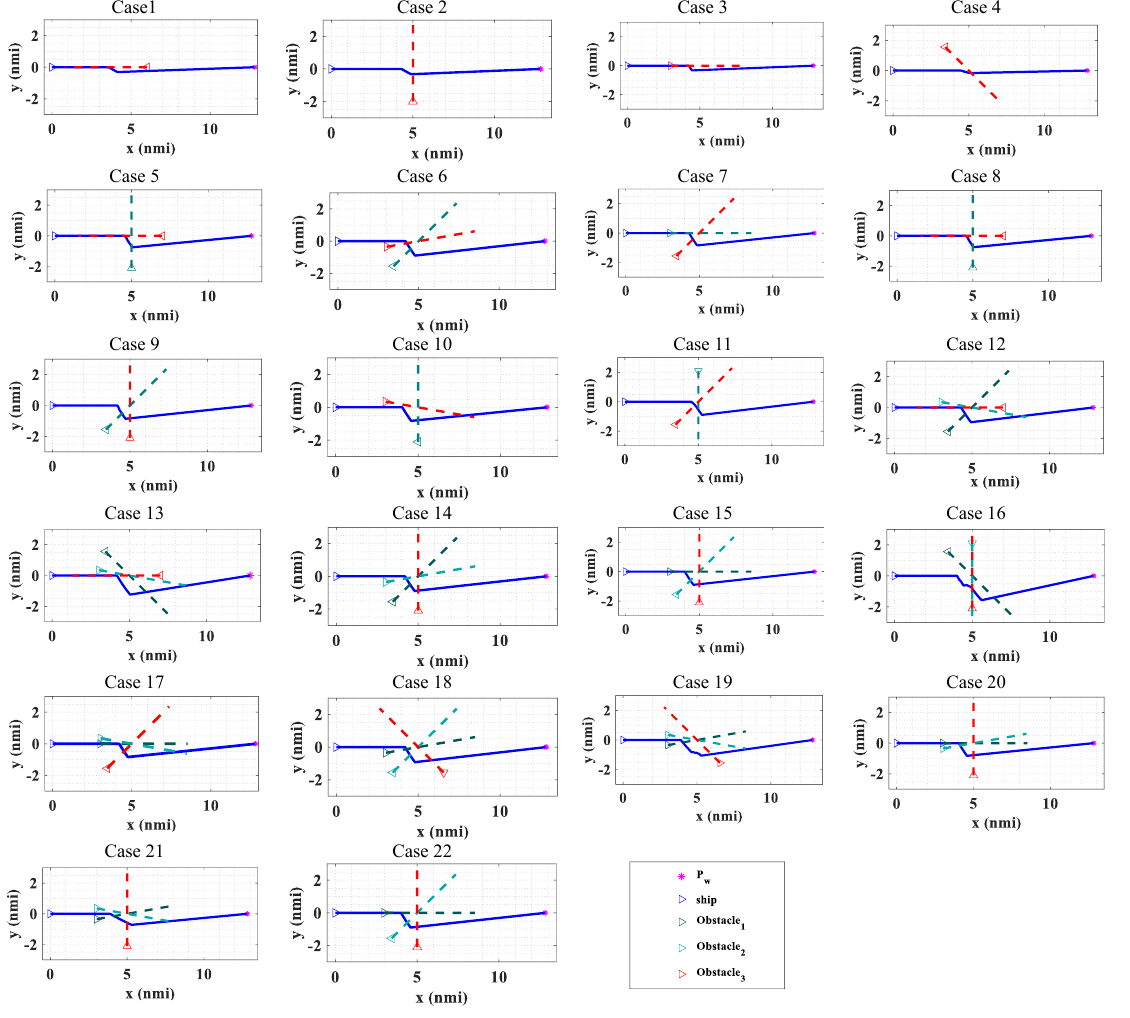} 	
\caption{Trajectories of the own ship and dynamic obstacles in all Imazu cases. The gifs of these Imazu cases can be found at \url{https://qubstudentcloud-my.sharepoint.com/:f:/g/personal/3057524_ads_qub_ac_uk/Euja59STLvBMkmxTNVd4DF8BLYHYDMb0HLOlXZsQat_y0g?e=sGcJzJ}}
	\label{fim}
\end{figure}

The simulation results of all cases are in Figure \ref{fim}. This figure shows the trajectories corresponding to own ship and moving obstacles in all the cases, which are represented by different colors. 
The simulation results for cases 1-3 shows that the ship is able to encounter the head-on, crossing, and overtaking situations well in advance according to the COLREG rules. In case 4, according to Rule 15, the target ship has to take action, but in this case, target ship does not act in time, and thus own ship successfully avoids it keeping a safe distance. In two (cases 5-11) and three (cases 12-22) moving obstacles scenarios, if there is a situation where the own ship is on the starboard side of the target ship, then the target ship is expected to make a maneuvering decision, however if target ship does not act in time, then own ship takes collision avoidance action. On the other hand, if the target ship is on the starboard side of the own ship, then own ship follows the COLREG's  and safely avoids the target ship.

\begin{table}[h!]
\fontsize{5.5}{6.6}\selectfont
\caption{Initial conditions}
\begin{center}
\label{tb1}
\begin{tabular}{ |c| c| c| c|}
\hline
 Imazu Cases & Obstacle 1 & Obstacle 2 & Obstacle 3\\ 
 \hline
 Case 1& $[6,0]^{\top}, 180^{\circ}$ & - & - \\  
 \hline
 Case 2 & $[5,-2]^{\top}, 90^{\circ}$ & - & - \\  
 \hline
 Case 3 & $[3,0]^{\top}, 0^{\circ}$ & - & - \\  
 \hline
 Case 4 & $[3.44,1.55]^{\top}, 255^{\circ}$ & - & - \\  
 \hline
 Case 5 & $[5,-2.1]^{\top}, 90^{\circ}$ &$[7,0]^{\top}, 180^{\circ}$  & - \\
 \hline
  Case 6 & $[3.4,-1.5]^{\top}, 45^{\circ}$ &$[3,-0.35]^{\top}, 10^{\circ}$  & - \\
 \hline
 Case 7 & $[3,0]^{\top}, 0^{\circ}$ &$[3.4,-1.5]^{\top}, 45^{\circ}$  & - \\
 \hline
 Case 8 & $[5,-2.1]^{\top}, 19^{\circ}$ &$[7,0]^{\top}, 180^{\circ}$  & - \\
 \hline
  Case 9 & $[3.4,-1.5]^{\top}, 45^{\circ}$ &$[5,-2.1]^{\top}, 19^{\circ}$  & - \\
  \hline
 Case 10 & $[3,0.3]^{\top}, -10^{\circ}$ &$[5,-2.1]^{\top}, 90^{\circ}$  & - \\
  \hline
Case 11 & $[5,2.1]^{\top}, -90^{\circ}$ &$[3.4,-1.5]^{\top}, 45^{\circ}$  & - \\
  \hline
  Case 12 & $[7,0]^{\top}, 180^{\circ}$ &$[3,0.3]^{\top}, -10^{\circ}$  &  $[3.44,-1.55]^{\top}, 45^{\circ}$ \\
  \hline
  Case 13 & $[7,0]^{\top}, 180^{\circ}$ &$[3,0.3]^{\top}, -10^{\circ}$  & $[3.4,1.5]^{\top}, 255^{\circ}$ \\
  \hline
  Case 14 & $[3.4,-1.5]^{\top}, 45^{\circ}$ &$[3,-0.3]^{\top}, 10^{\circ}$  & $[5,-2.1]^{\top}, 90^{\circ}$ \\
  \hline
   Case  15 & $[3,0]^{\top}, 0^{\circ}$ &$[-3.4,-1.5]^{\top}, 45^{\circ}$  & $[5,-2.1]^{\top}, 90^{\circ}$ \\
  \hline
  Case 16 & $[3.4,1.5]^{\top}, -45^{\circ}$ &$[5,2.1]^{\top}, -90^{\circ}$  & $[5,-2.1]^{\top}, 90^{\circ}$ \\
  \hline
   Case 17 & $[3,0]^{\top}, 0^{\circ}$ &$[3,0.3]^{\top}, -10^{\circ}$  & $[3.4,-1.5]^{\top}, 45^{\circ}$ \\
  \hline
   Case 18 & $[0.3,-0.3]^{\top}, 10^{\circ}$ &$[3.4,-1.5]^{\top}, 45^{\circ}$  & $[6.5,-1.5]^{\top}, 135^{\circ}$ \\
  \hline
  Case 19 & $[3,-0.3]^{\top}, 10^{\circ}$ &$[3,0.3]^{\top}, -10^{\circ}$  & $[6.5,-1.5]^{\top}, 135^{\circ}$ \\
  \hline
   Case 20 & $[3,0]^{\top}, 0^{\circ}$ &$[3,-0.3]^{\top}, 10^{\circ}$  & $[5,-2.1]^{\top}, 90^{\circ}$ \\
  \hline
  Case 21 & $[3,-0.3]^{\top}, 10^{\circ}$ &$[3,0.3]^{\top}, -10^{\circ}$  & $[5,-2.1]^{\top}, 90^{\circ}$ \\
  \hline
  Case 22 & $[3,0]^{\top}, 0^{\circ}$ &$[3.44,-1.55]^{\top}, 45^{\circ}$  & $[5,-2.1]^{\top}, 90^{\circ}$ \\
  \hline
\end{tabular}
\end{center}
\end{table}
\subsection{Comparative Analysis}
\begin{figure}[h!]
\centering
\includegraphics[width=\textwidth]{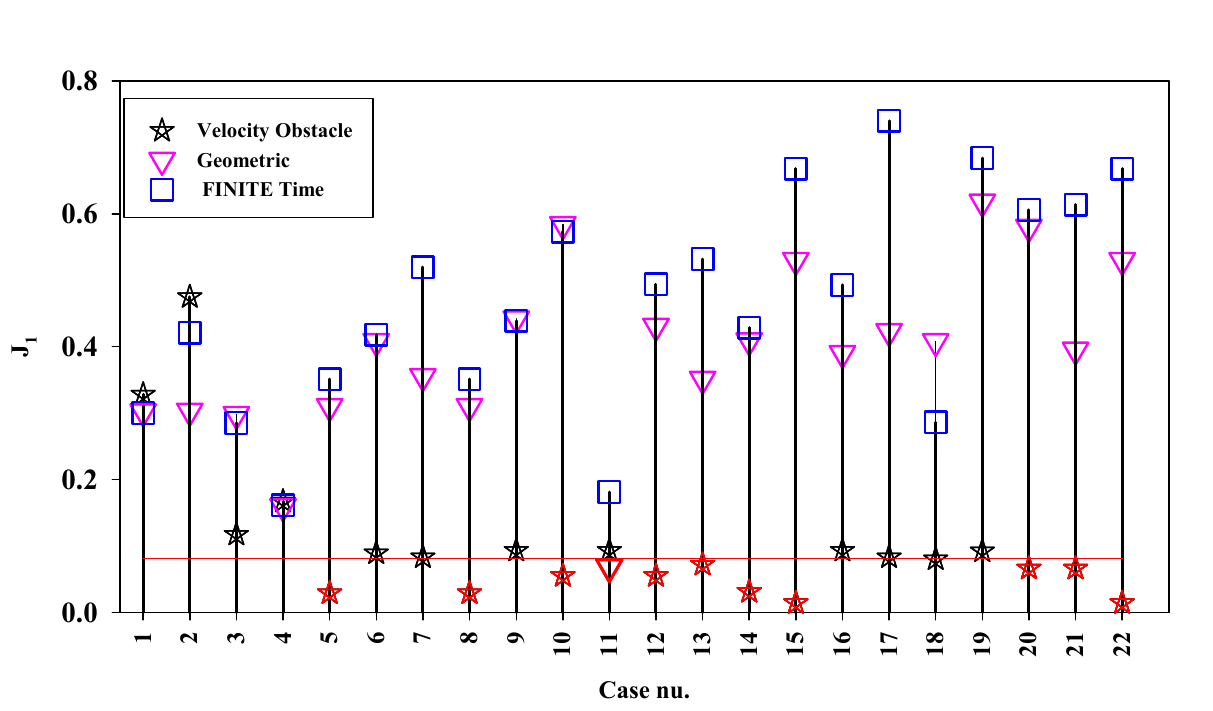} 	
\caption{Comparison of proposed method with other mentioned methods in terms of minimum obstacle distance}
	\label{cost1}
\end{figure}

\begin{figure}[h!]
\centering
\includegraphics[width=\textwidth]{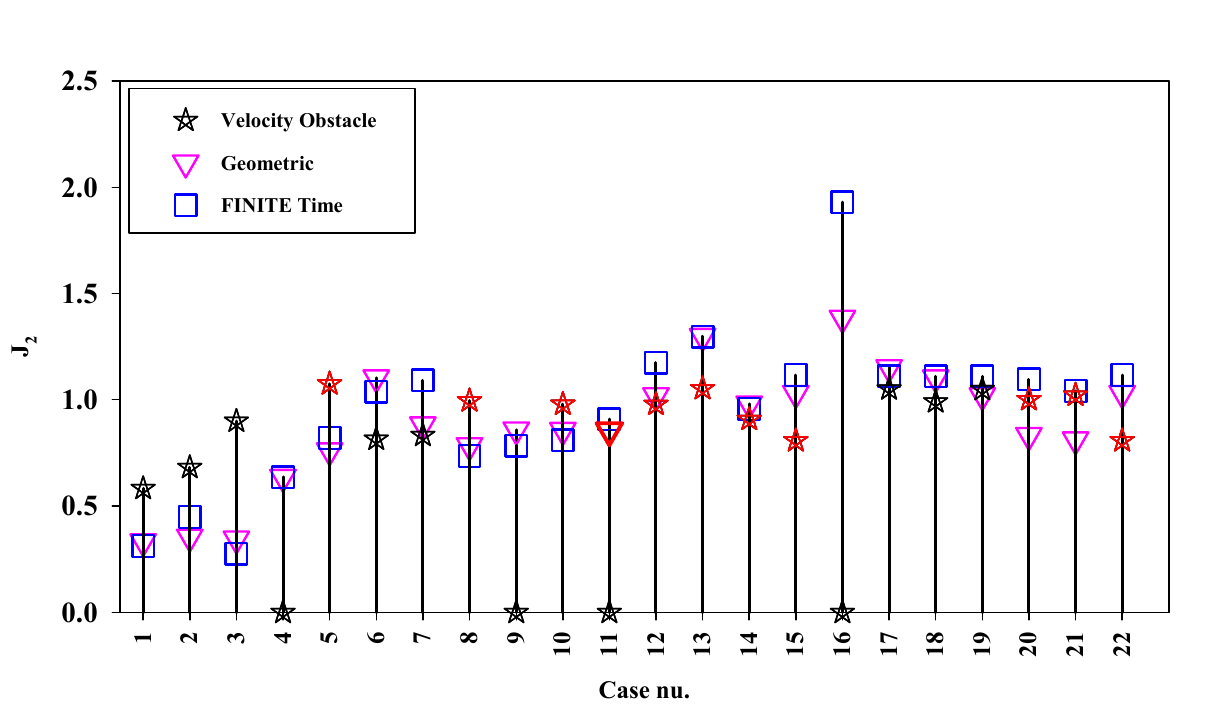} 	
\caption{Comparison in terms of maximum cross track error}
	\label{Cost2}
\end{figure}

\begin{figure}[h!]
\centering
\includegraphics[width=\textwidth]{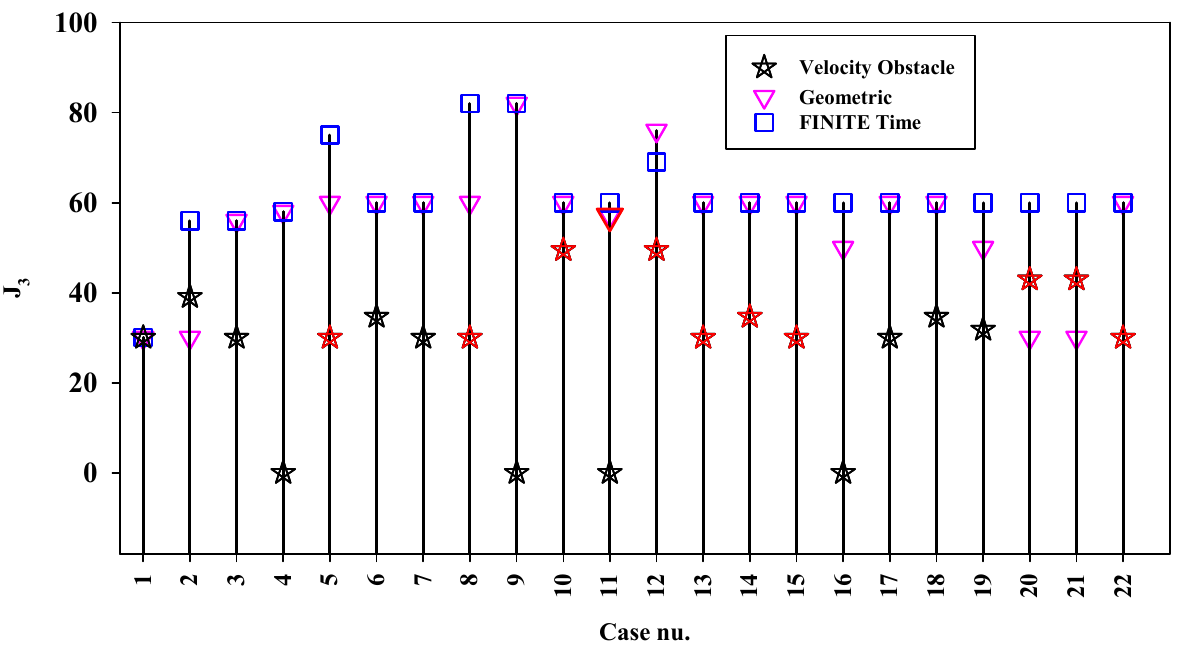} 	
\caption{Comparison in terms of control effort}
	\label{cost3}
\end{figure}

\begin{figure}[h!]
\centering
\includegraphics[width=\textwidth]{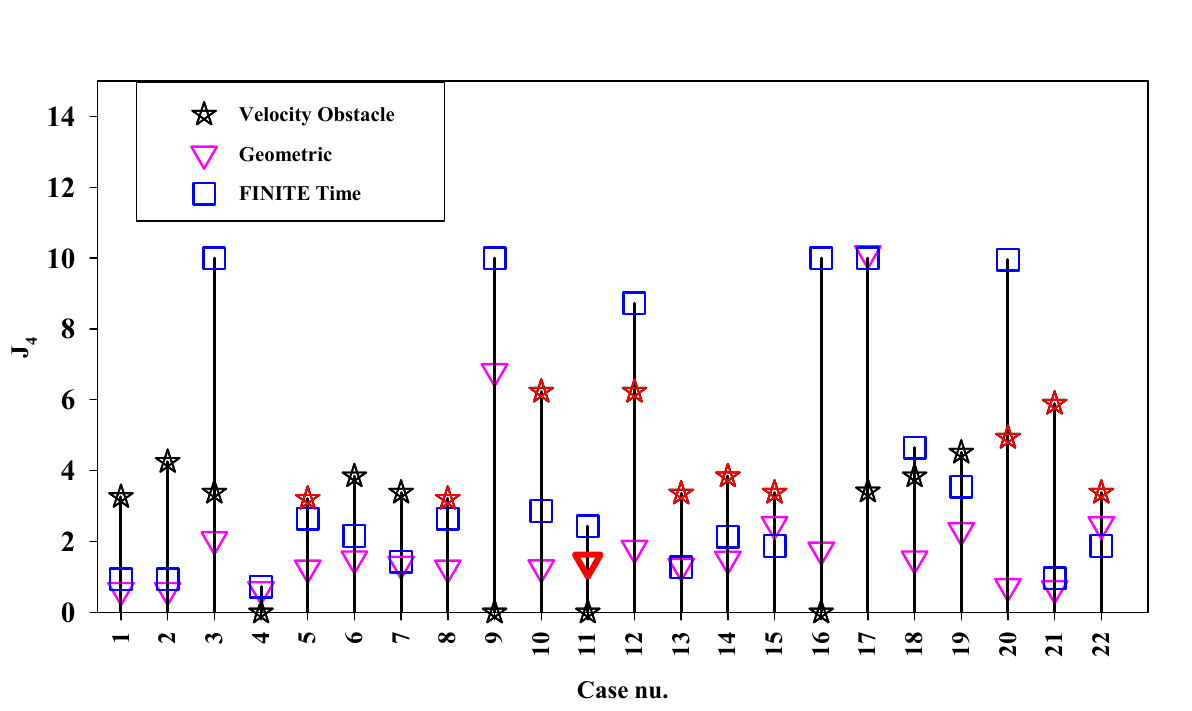} 	
\caption{Comparison in terms of rate of control effort}
	\label{cost4}
\end{figure}

To illustrate the effectiveness of our methodology, we conduct a comparative analysis between the proposed method and the classical Velocity Obstacle (VO) algorithm \cite{douthwaite2019velocity}.  The VO algorithm was originally designed for single obstacle avoidance, however, it exhibits a higher probability of failure when dealing with multiple obstacles, aligning with our expectations. We  also juxtapose our method with the Geometric Path-Planning Algorithm 
 \cite{dastgerdi2023geometric}.
 For the purpose of this comparison, we define six distinct metric functions that serve as key indicators of the COLAV algorithm's performance.
\begin{align}
J_1 &= \min_{i=1, \cdots, M}(\|p(t)-p_{mi}(t)\|),\\
J_2 &= \max(|\psi(t)-u(t)|), \\
J_3 &= \max|\Delta u(t)|, \\
J_4 &= \int_{0}^{T_F} |\Delta \dot{u}(\tau)|\,~d\tau, \\
J_5 &= T_F,\\
J_6 &= \int_{0}^{T_F} \sqrt{\left(\dot{x}(\tau)\right)^2 + \left(\dot{y}(\tau)\right)^2} d\tau.
\end{align}
 In Figure \ref{cost1}, we present the minimum obstacle distances, denoted as $J_1$, measured in nautical miles for all Imazu cases. The red line in this Figure represents $0.018$ nmi line, which is the safe distance $C_s$ for all scenarios.  
 Figure \ref{Cost2} showcases the maximum cross-track error, represented as $J_2$ in nautical miles. 
Figure \ref{cost3} depicts the maximum change in heading command angle, denoted as $J_3$ in degrees, for collision avoidance. 
In Figure \ref{cost4}, we visualize the variation in the rate of change of command heading angles over time, measured in degrees per second (deg/sec).

\begin{figure}[h!]
\centering
\includegraphics[width=\textwidth]{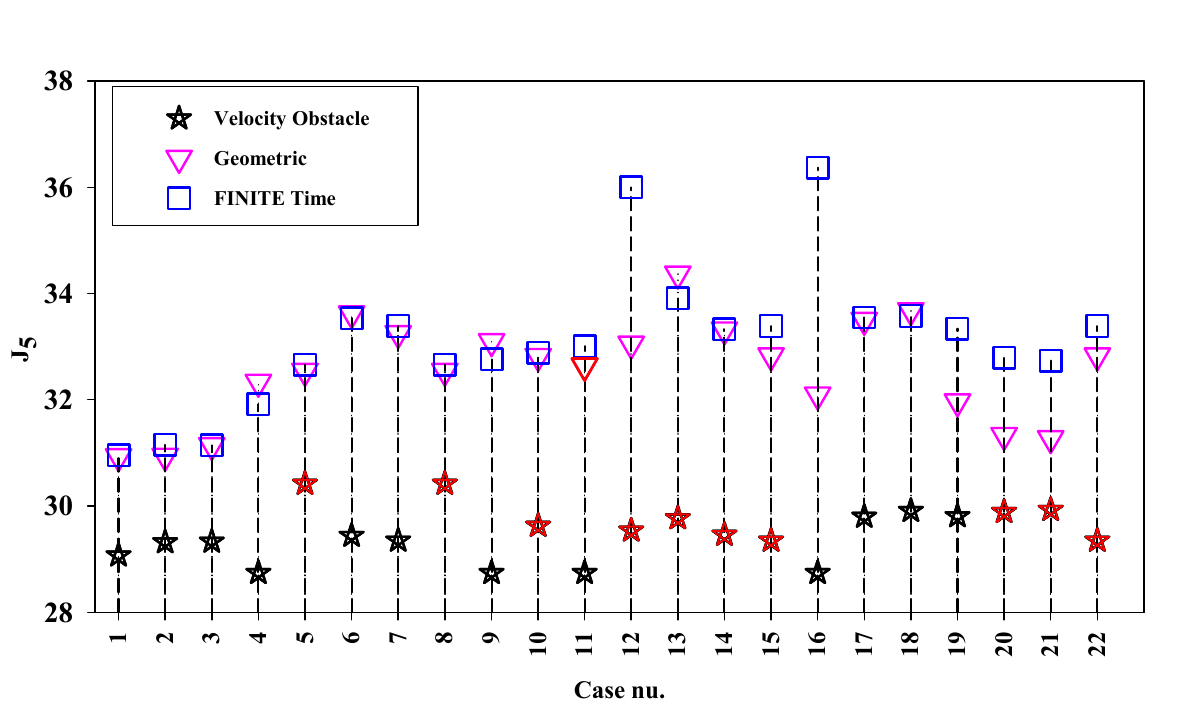} 	
\caption{Comparison in terms of final reaching time}
	\label{cost5}
\end{figure}

Finally, Figure \ref{cost5} illustrates the time taken to reach the waypoint, represented as $J_5 = t_f$ in minutes. 
We note for all Figures the failure of the algorithm in maintaining a safe distance with red color. 
According to Figure \ref{cost1}, it is evident that the minimum obstacle distance achieved by the proposed framework is larger than the  two algorithms in comparison. In Figure \ref{Cost2}, we observe that the maximum cross-track error of the proposed algorithm closely approximates that of the other algorithms. This implies that, in order to avoid obstacles while maintaining a substantial distance, we do not impose excessive penalties. Notably, in some cases, such as Case 1-3 of Imazu, the proposed algorithm outperforms the classical Velocity Obstacle algorithm, which is designed for scenarios involving a single obstacle.

\begin{figure}[h!]
\centering
\includegraphics[width=\textwidth]{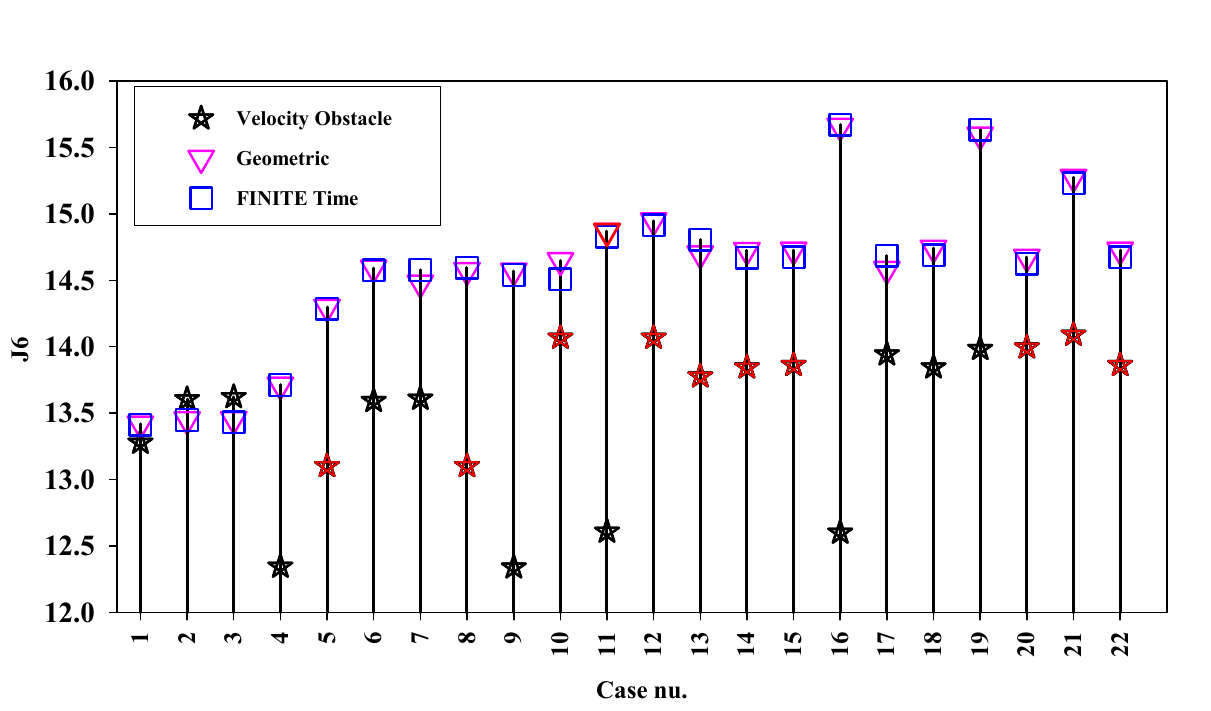} 	
\caption{Comparison in terms of path length traversed by the ship in final reaching time}
	\label{cost6}
\end{figure}

Figure \ref{cost3} illustrates the variations in the command heading angle, which are slightly higher when using the proposed algorithm. This outcome aligns with our expectations; when aiming to avoid obstacles with a significant distance while minimizing cross-track error, a greater variation in the command heading angle is often necessary. Figure \ref{cost4} and Figure \ref{cost5} depict the variation in heading angle over time and the time taken to reach the waypoint when utilizing the proposed algorithm.  Figure \ref{cost6} shows the path length traversed by the ship in the final reaching time.
In these aspects, the proposed algorithm demonstrates competitiveness with the other two methods.
As a last important comment, we note that our proposed framework is always safe, while failure to maintain the safe distance from the obstacles happens for the VO algorithm in cases 5, 8, 10, 12-15, 20-22, and the Geometric LOS algorithm in case 11.


\section{Conclusion}
Guidance and control systems are critically important for the overall performance and safety of the marine crafts. A new guidance law is introduced for waypoint navigation and collision avoidance in cascade with a new predefined-time heading autopilot. A predefined-time control is proposed for a heading autopilot design that is accurate and fast in tight turning maneuvers and Lyapunov methods are used to derive this control law. The equilibrium point of the heading autopilot error dynamics is proven to be predefined-time stable. It is guaranteed that ship reaches the waypoint in finite time with the bounded control with the proper illustration of the safe state space for multiple static obstacles and a moving obstacle, while safety is guaranteed also for moving dynamic obstacles under a mild assumption. The performance of our proposed framework is illustrated on a host of difficult realistic scenarios namely, Imazu problems. 

\begin{appendices}
\section{Proofs}
\subsection{Proof of Theorem \ref{the1}}
(i) The heading autopilot error dynamics (\ref{err}) with the proposed control $u$ (\ref{con1}) for $t_0\leq t< t_p$ becomes
\begin{align}\label{le}
   \dot{e}= -\frac{\eta(\text{e}^{e}-1)}{\text{e}^{e}(t_p-t)}. 
\end{align}
Consider the following candidate Lyapunov function for (\ref{le})
\begin{align}\label{ly}
V_e(e)=\frac{1}{2}e^2. 
\end{align}
The derivative of (\ref{ly}) along dynamics (\ref{le}) for $t_0\leq t< t_p$ is
\begin{align*}
\dot{V}_e=&~~e\dot{e}=-\frac{\eta e(\text{e}^{e}-1)}{\text{e}^{e}(t_p-t)}
\leq ~-\frac{\eta |e|(\text{e}^{|e|}-1)}{\text{e}^{|e|}(t_p-t)}.
\end{align*}
From (\ref{ly}), it can be inferred that $V_e\leq e^2$, which further results in $\sqrt{V_e}\leq |e|$. Thus, 
$$\dot{V}_e~\leq~-\frac{\eta \sqrt{V_e}(\text{e}^{\sqrt{V_e}}-1)}{\text{e}^{\sqrt{V_e}}(t_p-t)}.$$
Let $\zeta=\sqrt{V_e}$. Then, time derivative is $\dot{\zeta}=\frac{\dot{V}_e}{2\sqrt{V}_e}\leq -\frac{\eta (\text{e}^{\sqrt{V_e}}-1)}{2\text{e}^{\sqrt{V_e}}(t_p-t)}$ as long as $V_e(e(t))>0$,  which leads to  $\dot{\zeta}\leq-\frac{\eta (\text{e}^{\zeta}-1)}{2\text{e}^{\zeta}(t_p-t)}$. For $t\geq t_p$, $ \dot{V}_e=0$ with the control $u$ in (\ref{con1}).  Hence, from Lemma \ref{lem1}, the origin of the heading autopilot error dynamics is predefined-time stable and thus the ship heading angle converges to the desired guidance in a time $t_p$ prescribed in advance.

(ii) We show that the ship reaches the waypoint in finite time $T_F$. From (i), $\psi=\psi_{dg}$ for all $t\geq t_p\geq t_0$, (\ref{mar}) becomes 
\begin{align}\label{se}
\begin{split}
    \dot{x}=&~v\cos(\psi_{dg}),\\
    \dot{y}=&~v\sin(\psi_{dg}).
    \end{split}
\end{align}
Consider the distance between the ship and the waypoint, $d_w=\|p-p_w\|$. We choose Lyapunov function for (\ref{se}) as, 
\begin{align}\label{ly13}
V_w=\frac{1}{2}d_w^2.
\end{align}
The derivative of (\ref{ly13}) along the system trajectories (\ref{se}) for all $t\geq t_p\geq t_0$ is 
\begin{align*}
\dot{V}_w=&-v(x_w-x)\cos(\psi_{dg})-v(y_w-y)\sin(\psi_{dg}).
\end{align*}
We observe that
\begin{align}\label{er}
\begin{split}
\sin(\tan^{-1}(\frac{y_w-y}{x_w-x}))=&~\frac{(y_w-y)}{\sqrt{(x_w-x)^2+(y_w-y)^2}}\\ ~\cos(\tan^{-1}(\frac{y_w-y}{x_w-x}))=&~\frac{(x_w-x)}{\sqrt{(x_w-x)^2+(y_w-y)^2}},
\end{split}
\end{align}

  since $\sin(\tan^{-1}(x))=\frac{x}{\sqrt{1+x^2}}$ and $\cos(\tan^{-1}(x))=\frac{1}{\sqrt{1+x^2}}$. Then $\dot{V}_w$ becomes
\begin{align*}
\dot{V}_w=&~-\frac{v(x_w-x)^2}{\sqrt{(x_w-x)^2+(y_w-y)^2}}\\&~-\frac{v(y_w-y)^2}{\sqrt{(x_w-x)^2+(y_w-y)^2}}=-vd_w.
\end{align*}
Consequently, $$\dot{V}_w= -\sqrt{2}vV_w^{1/2}. $$
Hence, from Lemma \ref{lem2}, the system (\ref{se}) is finite-time stable and the ship reaches the waypoint in finite time $T_F=T_p+T_f+t_0$, with 
$T_f=~\frac{\sqrt{2}V_w(p(t_p))}{v}.$\\
(iii) We observe that maximum time from $t_0$ until $\psi=\psi_{dg}$ is predefined time $t_p$. During this time, the maximum distance traversed by the ship in time $t_p$ with velocity $v$ towards the waypoint, i.e., the \textit{maximum transient distance} is bounded by
\begin{align}\label{de}
    \cB_2(p(t_0), vt_p)=\{p:~\|p(t_p)-p(t_0)\|\leq vt_p\}.
\end{align}
Thus $p(t)\in \cB(p(t_0), vt_p),~\forall~ t\in[t_0, t_p]$. Moreover, we note

\begin{small}
	\begin{align}\label{ds}
 \begin{split}
	\dot{\psi}_{dg}=&~\frac{1}{1+(\frac{y_w-y}{x_w-x})^2}\frac{-v(x_w-x)\sin(\psi)+v(y_w-y)\cos(\psi)}{(x_w-x)^2}\\
	=&~\frac{-v(x_w-x)\sin(\psi)+v(y_w-y)\cos(\psi)}{(x_w-x)^2+(y_w-y)^2}.
 \end{split}
	\end{align}
		\end{small}
		From (\ref{ds}), we observe that $\dot{\psi}_{dg}$ becomes unbounded when the squared distance of the waypoint  from the ship: $d_w^2=\|p-p_w\|^2$ is very small.
For any $\delta>0$ such that $|x_w-x|\geq \delta$, $|y_w-y|\geq \delta$, and consequently, $(x_w-x)^2+(y_w-y)^2\geq \delta^2$. Then, $\dot{\psi}_{dg}$ is bounded by 
\begin{align*}
		|\dot{\psi}_{dg}|
		\leq &~\frac{v(|x_w-x|)|\sin\psi|+v(|y_w-y|)|\cos(\psi)|}{(x_w-x)^2+(y_w-y)^2}\\
		\leq&~\frac{v\delta+v\delta}{\delta^2}\leq \frac{2v}{\delta}.
		\end{align*}
		
 To find the minimum distance $\delta$ between the ship and the waypoint such that $u(t)\in\mathcal{U}$ at all times as defined in (\ref{con1}).
Given $t_p>0$, we enforce
\begin{align}\label{ww}
\begin{split}
|u|\leq &~m,\\
\implies \frac{1}{a}|\dot{\psi}_{dg}|+|\psi_{dg}|+\frac{\eta |1-\text{e}^{-e}|}{a|(t_p-t_0)|}\leq&~m,\\
\implies\frac{2v}{a\delta}+\pi+\frac{\eta |1-\text{e}^{-e}|}{a|(t_p-t_0)|}\leq&~ m,
\end{split}
\end{align}
or, when (\ref{no}) holds.
 Thus, $p(t_0)\in \bar{\cB}_2(p_w, \delta)\cup\bar{\cB}_2(p(t_0), vt_p)$ 
 holds for all $t\geq t_0$ and $u(t)\in\mathcal{U}$.
\hfill $\blacksquare$
\subsection{Proof of Theorem \ref{thee2}}
By Assumption 1, the ship starts at time $t_0$ with the guidance law $\psi_{dg}(p, p_w)$ in (\ref{ni}) since $G_{11}$ holds. Consequently, by Theorem \ref{the1}, $\psi(t)=\psi_{dg},$  $t\geq t_p+t_0$, and convergence to waypoint $p_w$ terminal set $\cB_2(p_w, \delta)$ is guaranteed in finite time $T_F$ until ${G}_{11} \land \bar{G}_{12}$ holds. 

We denote the time when the conditions (\ref{cc1})-(\ref{df}) are fulfilled by $T_1>t_p>t_0$. At this time, the ship enters the obstacle avoidance mode. Since the controller (\ref{con1}) with the guidance law $\psi_{dg}(p, V_1)$ in (\ref{ni}) is applied to system (\ref{mar}) until the condition  $L_1\land L_2$ from the control (\ref{cont}) is satisfied, the  heading angle is $\psi(t)=\psi_{dg}(p, V_1)$ for all $t\geq T_1+ t_p+t_0$  from Theorem \ref{the1}. Also, it is ensured that $p(t)\in \cB_2(V_1, \delta)$ in time $t\leq T_{F1}$, where  $ T_{F1}>T_1+t_p+t_0$, is  finite time. 

Moreover, in the obstacle avoidance mode, we observe that the maximum time where the conditions (\ref{cc1})-(\ref{df}) are fulfilled,  until $\psi=\psi_{dg}(p, V_1)$, is  $T_1+t_p$. During this time, the  distance traversed by the ship in time $t_p$ with velocity $v$ is bounded by $vt_p$.

To ensure safety,  $d_s(T_1+t_p+t_0)\geq ~C_s$.
 Thus, with the switching condition, $d_s(t)\leq d_{safe}=C_s+vt_p$, it is ensured that the ship heading is equal to the desired heading provided from the  obstacle avoidance law, before the ship comes at a distance $C_s$ from the obstacle. 
After $t\geq T_{F1}$, until the condition  $\bar{L}_1\lor \bar{L}_2$ from the control (\ref{cont}) is satisfied,  the ship remains at a distance greater than $C_s$ from the obstacle. 
We denote the time when  $\bar{G}_{11}$ is satisfied by $T_2$ such that $T_2>T_{F1}$, the ship exits the avoidance mode and resumes to the waypoint reaching mode  towards $p_w$ (goal).  

At time $T_2$, ship remains at a safe distance greater than or equal to $C_s$ from the obstacle $p_s$ since there is no intersection of waypoint LOS set $\cF(p(T_2))$ with the unsafe set $\cB_{\infty}(p_s, C_s)$ from the satisfaction of the condition $\bar{G}_{11}$ as this unsafe set comprises of the obstacle $p_s$ inflated by the safe distance $C_s$. Since, the controller (\ref{con1}) with the guidance law $\psi_{dg}(p, p_w)$ (\ref{ni}), is applied to (\ref{mar}), the vehicle heading angle is $\psi(t)=\psi_{dg}(p, p_w)$ for all $t\geq T_2+ t_p+t_0$.  Thus,  the ship remains at a distance greater than safe distance $C_s$ from the obstacle position for all time $t\geq T_2$.

Last, it is ensured from Theorem \ref{the1} that $p(t)\in \cB_2(p_w, \delta)$ in time $t\leq T_{F}$, where  $ T_{F}>T_2+t_p+t_0$ is a finite interval.  Hence, it follows that ship reaches the waypoint $p_w$ terminal set $\cB_2(p_w, \delta)$ while $p(t)\notin \cB_{\infty}(p_s, C_s) ~\text{for all time}~ t\in[t_0, T_F].$ \hfill $\blacksquare$
\\
\subsection{Proof of Theorem \ref{thee3}}

 (i)  By Assumption 1, the ship starts at time $t_0$ with the guidance law $\psi_{dg}(p, p_w)$ in (\ref{ni}) since $G_{11}$ holds.  Thus, from Theorem \ref{thee2}, the ship reaches  $\cB_2(p_w, \delta)$ in finite time $T_F$ as long as condition ${G}_{11} \land \bar{G}_{12}$ holds. \\ 
 
If this is not the case, as in Theorem \ref{thee2}, we denote the time when the conditions (\ref{cc1})-(\ref{df}) are fulfilled, with $T_1>t_p>t_0$. It is then ensured that $p(t)\in \cB_{2}(V_1,\delta)$ in time $t\leq T_{F1}$, where $T_{F1}=T_1+\text{TCPA}$ according to the distance $d_2$ (\ref{fv}) since we consider the static virtual waypoint $V_1=V_1(T_1)$. 
Consider the hyperplance $H$ where the trajectory of the dynamic obstacle $(x_m, y_m)\in\mathbb{R}^2$ lies, satisfying the equation \begin{align}\label{dev}
  H:~~~   a_1x_m+b_1y_m+d_1=0.
 \end{align}
 with $a_1, b_1$, $d_1$ constants.
 At time $T_{F1}$, the projection of the ship position $p(T_{F1})$ on $H$ (\ref{dev}) is on  the obstacle position $p_m(T_{F1})$ or in front it, since the time taken by the obstacle $p_m$ to reach $p_{mc}$ is $T_1+TCPA=T_{F1}$.
 After $t\geq T_{F1}$, the ship $p(t)$ and waypoint $p_w$ are on the same side of the line (\ref{dev}) for the cases shown in Figure \ref{f213}, \ref{f213r} (i)-(ii). Thus, there always exists the time $T_2\geq T_{F1}$ when  $G_{23}$ is satisfied and the ship exits the avoidance mode and resumes to the waypoint reaching mode  towards $p_w$ (goal), reaching it in time $t\leq T_F$.  However, for the case shown in Figure \ref{f213r} (iii), after  $t\geq T_{F1}$, the ship $p(t)$ and the waypoint $p_w$ are separated by $H$ (\ref{dev}).  To reach the waypoint in finite time $T_F$, the ship has to cross this line. In this configuration, there are two possible outcomes:  
 
 (a) the collision avoidance conditions (\ref{cc1})-(\ref{df}) are triggered one more time at $t=T_{11}$. In this case, the new CPA $p_{mc}(T_{11})$ and the set $Z_s(T_{11}),~ V_1=V_1(T_{11})$ are calculated again, while it is ensured again from Theorem \ref{thee2} that $p(t)\in \cB_2(V_1, \delta)$ in finite time $T_{F2}$.  We note that after time $T_{F2}$, collision avoidance conditions cannot be triggered again as the trajectory of the ship $p(t)$ with the constant heading angle and the trajectory of the moving obstacle are diverging. Indeed, after $T_{F2}$,  control (\ref{cont}) remains constant until  $G_{23}$ is satisfied.
 
 (b) the ship resumes to the waypoint reaching mode, i.e., $G_{23}$ is satisfied.

\begin{figure}[H]
	\centering
	\includegraphics[ width=8.8cm]{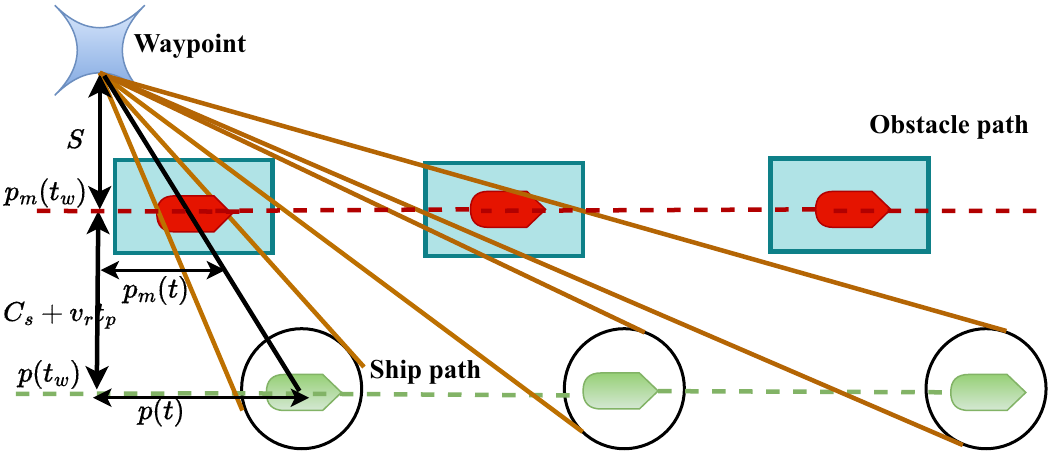} 	\caption{Possible motion of the dynamic obstacle $p_m$ and ship $p_s$ (parallel trajectories}
	\label{f213pr}
\end{figure}

\begin{figure}[H]
	\centering
	\includegraphics[height=2.1in, width=8.8cm]{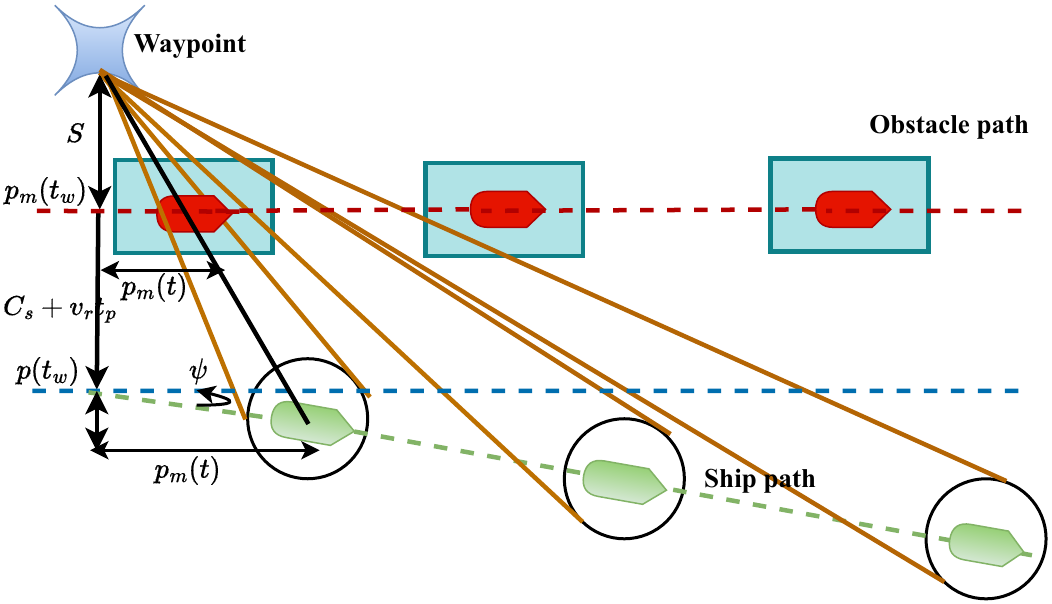} 	\caption{Possible motion of the dynamic obstacle $p_m$ and ship $p_s$ (not parallel trajectories).}
	\label{f213pr2}
\end{figure}

What only remains to show is that in case (a), there is a finite time $T_2$ so that $G_{23}$ is triggered. 
Indeed,  $G_{23}$  is satisfied at some time $T_2$ when there is no intersection of $\cF(p(T_2))$ with the reachable set $\mathcal{F}_0(p_m(T_2),  T)$. We visualize 
ship and obstacle paths in Figures  \ref{f213pr} and \ref{f213pr2} while determining the waypoint LOS. We show by contradiction that we  achieve clear LOS in finite time. To this purpose, let us suppose that the LOS is always obstructed  for all times. Then, we have, from Figure \ref{f213pr}, 

\hfill 
\begin{align}\label{ff}
\begin{split}
    \frac{x_m}{S}=&~\frac{x}{S+C_s+vt_p} \Rightarrow \\
    \frac{v_o(t-t_w)}{S}=&~\frac{v(t-t_w)}{S+C_s+vt_p} \Rightarrow \\
    \frac{v}{v_o}=&~\Big(\frac{1+\frac{C_s}{S}-\frac{v_ot_p}{S}}{1-\frac{v_ot_p}{S_1}}\Big),
    \end{split}
\end{align}
where $t_w\geq T_1+T_c$, which however by the statement of the Theorem cannot happen, thus a contradiction is reached. For the other case shown in Figure \ref{f213pr2}, when the trajectories of the two ships are not parallel,  it holds
\begin{align}\label{ff4}
\begin{split}
    \frac{x_m}{S}=&~\frac{x}{S+C_s+vt_p+|y-y(t_w)|}\Rightarrow \\
    \frac{v_o(t-t_w)}{S}=&~\frac{v\cos\psi (t-t_w)}{S+C_s+vt_p+v\sin \psi (t-t_w)},
    \end{split}
\end{align}
which is not possible for all $t$. Thus, if the ratio of velocities do not satisfy (\ref{ff}), then necessarily $v=\alpha v_o$,  for some  constant $\alpha\neq 1$, and consequently the distance between the ship and obstacle $\Delta p=|v-v_0|t$ increases arbitrarily with time. After sufficient time $t^{\star}$, condition $G_{23}:~~\cF_0(p_m(t^{\star}),T)\cap\cF(p(t^{\star})) =\emptyset$ is satisfied.  Hence the ship reaches  $\cB_2(p_w, \delta)$ in time $t\leq T_F$.

(ii) We need to show that $p(t)\notin \cB_{\infty}(p_m(t), C_s)$ for all time $t\in[t_0, T_F],$ when $G_{12}$ (\ref{df}) is satisfied, 
with  $d_{safe}=d_2+ L(T_1)+vt_p$. 
 At time $T_1$, the ship enters the obstacle avoidance mode and the distance between the dynamic obstacle and the vertex $V_i, ~i=1,2$ of set $\mathcal{Z}_s(T_1)$ is $d_{mv}=\|p_m(T_1)-V_1\|=\sqrt{2}C_s$ from the construction of the set $\mathcal{Z}_s(T_1)$.
In the obstacle avoidance mode, we observe that the maximum time where the conditions (\ref{cc1})-(\ref{df}) are fulfilled,  until $\psi=\psi_{dg}(p, V_1)$, is  $T_1+t_p$. During this time, the  distance traversed by the ship in time $t_p$ with velocity $v$ is bounded by $vt_p$. 
To ensure safety,  $d_m(T_1+t_p+\text{TCPA}+t_0)=\|p(T_1+t_p+\text{TCPA}+t_0)-p_m(T_1+t_p+\text{TCPA}+t_0)\|\geq ~C_s$.
 Thus, with the switching condition, $d_s(t)\leq d_{safe}=d_2+L(T_1)+vt_p$, it is ensured that the ship heading is equal to desired heading provided from the  obstacle avoidance law, before the ship comes at a distance $C_s$ from the obstacle $p_m$, since obstacle is at $p_{mc}(T_1)$ and $d_m(T_1+t_p+\text{TCPA}+t_0)=\sqrt{(L^2(T_1)+B^2(T_1))}\geq ~C_s$. 
 After $t\geq T_{F1}$,  for the cases shown in Figures \ref{f213}, \ref{f213r} (i)-(ii), the ship remains at a distance greater than $C_s$ from the obstacle $p_m$. For the case Figure \ref{f213r} (iii), the ship can again switch to the collision avoidance mode as described in (a) and hence in a  similar way, safety from the vertex of $\mathcal{Z}_s(T_{11})$ is ensured. At time $T_2$, the ship remains at a safe distance greater than or equal to $C_s$ from the obstacle $p_m$ since there is no intersection of waypoint LOS set $\cF(p(T_2))$ with the unsafe set $\cF_0(p_m(T_2), T)$ from the satisfaction of the condition $G_{23}$ as this unsafe set comprises of the obstacle trajectory inflated by the safe distance $C_s$. Since, the controller (\ref{con1}) with the guidance law $\psi_{dg}(p, p_w)$ (\ref{ni}), is applied to (\ref{mar}), the ship heading angle is $\psi(t)=\psi_{dg}(p, p_w)$ for all $t\geq T_2+ t_p+t_0$.  Thus,  the ship remains at a distance greater than safe distance $C_s$ from the obstacle position for all time $t\geq T_2$.
Hence, it follows that ship reaches  $\cB_2(p_w, \delta)$ while $p(t)\notin \cB_{\infty}(p_m, C_s) ~\text{for all time}~ t\in[t_0, T_F].$ \hfill  $\blacksquare$
\end{appendices}

\end{document}